\documentclass[12pt,preprint]{aastex}
\usepackage{longtable}

\begin{document}
\title{
Spatially Resolved Galaxy Star Formation and its Environmental Dependence II. Effect of the Morphology-Density Relation
}

\author{Niraj Welikala\altaffilmark{1}, Andrew J. Connolly\altaffilmark{2},
        Andrew M. Hopkins\altaffilmark{3}, Ryan Scranton\altaffilmark{4}
       }

\altaffiltext{1}{Laboratoire d'Astrophysique de Marseille, 38 Rue Fr{\'e}d{\'e}rique Joliot-Curie, 13388 Marseille Cedex 13, France, niraj.welikala@oamp.fr}
\altaffiltext{2}{Department of Astronomy, University of Washington, Box 351580, Seattle, WA 98195-1580,USA,
ajc@astro.washington.edu}
\altaffiltext{3}{Anglo-Australian Observatory, P.O. Box 296, Epping, NSW 1710, Australia,
ahopkins@aao.gov.au}
\altaffiltext{4}{Department of Physics, University of California at Davis, One Shields Avenue, Davis, CA 95616-8677, USA}

\begin{abstract}

In this second of a series of papers on spatially resolved star formation,
we investigate the impact of the density-morphology relation of galaxies
on the spatial variation of star formation (SF) and its dependence on
environment. We find that while a density-morphology relation is present for
the sample, it cannot solely explain the observed suppression of SF in galaxies
in high-density environments. We also find that early-type and late-type
galaxies exhibit distinct radial star formation rate (SFR) distributions,
with early-types having a SFR distribution that extends further
relative to the galaxy scale length, compared to late-types at all densities.
We find that a suppression of SF in the highest density environments is found
in the highest star forming galaxies for both galaxy types. This
suppression occurs in the innermost regions in late-types ($r\le 0.125$
Petrosian radii), and further out in radius in early-types ($0.125< r
\le 0.25$ Petrosian radii). When the full sample
is considered no clear suppression of SF is detected, indicating that the
environmental trends are driven only by the highest SF galaxies.
We demonstrate that the density-morphology relation alone cannot account
for the suppression of SF in the highest density environments. This points
to an environmentally-governed evolutionary mechanism that affects the SF
in the innermost regions in both early and late-type galaxies. We suggest
that this is a natural consequence of the ``downsizing" of SF in galaxies.
 
\end{abstract}


\section{Introduction
\label{sec:intro}}

The morphology of a galaxy is an indicator of its internal structure
and of the dynamical processes that give rise to it. Edwin Hubble
first classified galaxies according to their morphology along the
so-called `Tuning-fork diagram' \citep{Hub:36,San:61}. Some of the more common
classifications are based on visually determining the galaxy morphologies,
\citep{deV:91,Lin:08}, while there have been many attempts over the past
decade at quantifying morphology using measurements of concentration, color,
surface brightness profiles or features in the galaxy spectrum
\citep{Abr:03, Con:06,Got:03,Str:01}. Broadly speaking, galaxies classified
as ``early-type'' have morphologies that are typically elliptical and lenticular
while ``late-type'' morphologies are either spiral or irregular. Early-types
also tend to be redder, more luminous, more gas-poor and have older
stellar populations than do late-types.

In dense environments the galaxy population is dominated by
early-types. This was first established by \citet{Dre:80} who studied
55 nearby, rich clusters and found that the S0 fraction increases
steadily and the elliptical galaxy fraction increases sharply at the
highest densities while the fraction of spiral galaxies decreases
steadily with increasing local galaxy density in all clusters. This
density-morphology relation implies that the morphology of the galaxy is
shaped by the physical mechanisms that are prevalent in that particular
environment. The density-morphology relation was also observed in groups
of galaxies. \citet{Pos:84}, looking at the same dataset found that the
relation extended to galaxy group environments identified in the CfA
Redshift Survey. The relation is also observed in X-ray selected poor
groups \citep{Tra:01}. However \citet{Whi:95} found that the relation
is very weak or non-existent in groups.

More recently, \citet{Got:03} studied the density-morphology relation and the morphology-cluster-centric-radius relation in galaxies in the Sloan Digital Sky Survey (SDSS) and found that the elliptical fraction increases and the disc fraction decreases towards increasing galaxy density. They also found that there are two characteristic changes in both relations, suggesting that two different mechanisms are responsible for the relations. In the sparsest regions, they found that both relations became less noticeable, but in the intermediate-density regions, they found that the fraction of S0s increases in higher density environments, whereas the disc fraction decreases. 
In the densest regions, the S0 fraction decreases rapidly and the elliptical fraction increases, suggesting that a second mechanism is responsible for any morphological transformation of galaxies in the cluster cores. \citet{Par:07}, using SDSS galaxies, found that the fraction of early-type galaxies is a monotonically
increasing function of local galaxy density and luminosity.

The density-morphology relation has also been detected at higher
redshifts. \citet{Dre:97} found a strong relation for centrally
concentrated clusters at $z \approx 1$ but not for less concentrated
ones. \citet{Fas:00} studied nine clusters in the redshift range
$0.1 \le z \le 0.25$ and found a density-morphology relation in
high elliptical concentration clusters though not in low elliptical
concentration clusters, consistent with \citet{Dre:97}. They also traced
the morphological fraction as a function of cosmic time and, confirming
the ``Butcher-Oemler effect" \citep{BO:78}, found that the S0 fraction
decreases with increasing redshift while the spiral fraction increases.
However, \citet{Hol:07}, using galaxies in five massive X-ray clusters from $z=0.023$ to $z=0.83$ found that the evolution of the morphology-density relation differs considerably between galaxies selected by stellar mass and those selected by luminosity, the early-type fraction changing much less in mass-selected samples. The result is echoed by \citet{VanderWel:07} who used galaxies in the SDSS and the GOODS-South field and found little change in the morphology-density relation since $z \approx 0.8$ for galaxies more massive than $0.5 \,M_{\star}$. 

The same physical mechanisms that are proposed to explain the
relation between SFR and environment have been proposed to explain
the morphology-density relation as well. These include ram pressure
stripping of gas \citep{GG:72}, gravitational interactions between
galaxies \citep{BV:90}, galaxy harassment via high-speed encounters
\citep{Moo:96} and galaxy mergers. However, little observational
evidence exists to suggest that these processes drive the evolution in
galaxies. On the theoretical front, the combination of semi-analytic
models with N-body simulations of cluster formation has enabled the
density-morphology relation to be simulated. \citet{Dia:01} derived the
relation assuming that galaxy morphologies (determined in the simulation
using a bulge-to-disc ratio) in clusters are solely determined by
their merging histories. They found good agreement with data from the
CNOC1 sample \citep{Yee:96} for bulge-dominated galaxies. \citet{Ben:01}
found that a strong density-morphology relation was established at $z=1$
which was similar to that at $z=0$ but their results suggested that more
than one of the physical mechanisms mentioned above may have to be used
to explain the relation.

 In \citet{Wel:08} (hereafter Paper I), we studied the spatial variation of SF within galaxies as a function of the galaxy environment, for 44,964 galaxies in the SDSS. We showed that the star formation rate (SFR) in galaxies in high-density regions is suppressed compared to those in lower-density regions. We showed that this suppression occurs in the innermost regions within galaxies ($r\le 0.25 R_p$ where $R_p$ is the Petrosian radius). The study dealt with galaxies of all morphologies. We now aim to extend that investigation by focusing on the following three questions:

\begin{enumerate}
\item  Do early-type and late-type galaxies, which have distinct radial
light profiles, also have distinct radial distributions of SF? This may
be associated with different formation mechanisms or evolutionary
histories in either galaxy type.
\item Does any type-dependent change in the spatial distribution of SFR
occur for all galaxies of that type uniformly? Or is it restricted to a
sub-population, such as the highly star forming systems? In Paper I we
established that the suppression of SF seen in high density environments for
the full sample is a consequence of suppression only in the highly
star forming sub-population. We want to establish whether the trends in
early and late type galaxies separately are also dominated by this
active sub-population.
\item Is the observed environmental dependence of SFR in galaxies a
consequence of, or in addition to, the reduced average SFRs expected
simply from the higher proportion of early-types in high density environments?
\end{enumerate}

We describe our method and approach in \S\,\ref{sec:analysis}, and our
results in \S\,\ref{sec:results}. These are discussed in
\S\,\ref{sec:discussion}, where we explore an evolutionary explanation for
the observed trends. We present our conclusions in \S\,\ref{sec:conclusions}.
We assume throughout that $\Omega_{\Lambda}=0.7$, $\Omega_{\rm M}=0.3$,
and $H_{0}=75\,{\rm km\,s^{-1}\,Mpc^{-1}}$.

\section{Method}
\label{sec:analysis}

\subsection{The Pixel-z and Environmental Measures}

In Paper I we determined the radial variation of SF in 44,964 galaxies
in the Fourth Data Release (DR4, \cite{Ade:06}) of the Sloan Digital Sky Survey (SDSS, \cite{Yor:00}) using a technique called
`pixel-z' \citep{Wel:08,Con:03}. This technique involves fitting spectral energy distributions
(SEDs) generated from the stellar population synthesis models of
\citet{BC:03} to the photometric fluxes in five bands ($u,g,r,i,z$)
in individual pixels of each galaxy (see Paper I for details). The calculation of the 
fluxes in each pixel of every galaxy image is described in Appendix A. The SFR
in each pixel is calculated directly from the best fit SED to the pixel
fluxes and an associated uncertainty from that fit is also calculated.

Given the SFR determined for every pixel in every galaxy, we determined
both the total SFR in each galaxy and the spatial variation of SF within
galaxies in the sample. We used this to explore the environmental dependence
of the radial variation of SFR in galaxies. We found that the suppression of
total SFR in galaxies in high density regions is an effect seen most strongly
in the high-SF population and is a consequence of centrally suppressed SF.

In the current analysis we use the same galaxies in DR4, the same environmental measure based on
the spherical density estimator, and the same pixel-z estimates for SFR
as in Paper I. We retain the same two morphological classes as in Paper I,
as used also by \citet{Got:03}, based on discriminating galaxies according
to their inverse concentration index $C_{in}$ which is found to be correlated with the galaxy type \citep{Shi:01,Str:01}. We use $C_{in}$ to classify galaxies into two broad morphological classes: early-types (E,S0,Sa) and late-types (Sb, Sc and Irr). $C_{in}$ is defined as the
ratio of the radius containing 50\% of the Petrosian flux ({\tt petroR50\_r}) to the radius
containing 90\% of the Petrosian flux in the galaxy ({\tt petroR90\_r}). {\tt petroR50\_r}
and {\tt petroR90\_r} are obtained from the {\tt PhotoObjAll}
table in the Catalog Archive Server (CAS) in DR4\footnote{http://cas.sdss.org/dr4/en/}. A descriptions of how the Petrosian flux is calculated is given in \cite{Bla:01}.
The selection results in 27\,993 early-type galaxies
($C_{in} \le 0.4\,$) and 16\,971 late-type galaxies ($C_{in} > 0.4$) classified
according to this parameter.
 
To establish that our results are not sensitive to, or biased by, this
choice of morphology proxy, we duplicate our analysis using the Sersic
index. The Sersic model for the surface brightness in a galaxy is
given by $I(r) = I_0 e^{(-r/r_0)^{1/n}}$,
where $I(r)$ is the intensity at an angular radius $r$, $I_0$ is the
central intensity, $r_0$ is the characteristic radius, and $n$ the
Sersic index or profile shape parameter. An exponential profile is
recovered with $n = 1$, while $n = 4$ gives the traditional de Vaucouleurs
profile. The Sersic indices for the galaxies are found by cross-matching galaxies in our sample with those found in the NYU Value Added Galaxy Catalog (NYU-VAGC\footnote{http://sdss.physics.nyu.edu/vagc} and \cite{Bla:05}) where the Sersic profiles for the galaxies are made available. The Sersic profiles are stored in the {\tt sersic\_catalog.fits} files and the values of $n$ for the objects are stored in the column {\tt SERSIC\_N[i]} where {\em i} denotes a particular passband. Unlike concentration, the Sersic indices have been corrected for the effects of seeing. In our analysis we classify galaxies with $n<2$ as late-types and those with $n>2$ as early-types.

We first investigate the effect of an improved method to obtain the total SFRs in galaxies, based on the method for obtaining the mean radial SFR.

\subsection{The Total Galaxy SFR-Density Relation: An Improved Estimate}
\label{subsec:totsfr}

In Paper I, we investigated how the distribution of galaxy SFRs changes as a function
of the local (spherical) galaxy density. The SFR due to stellar populations within each pixel of a galaxy is assumed to follow an exponential functional form $\Psi(t) = \Psi_0 e^{(-t/\tau)}$. In Paper I, the total SFR for each galaxy was calculated from the weighted sum of the SFRs in each pixel in the galaxy image, where the weights correspond to the reciprocal of the square of the fractional error on the SFR in each pixel. The weighting is done to avoid giving undue significance to poorly constrained pixels, such as pixels dominated by the sky background, so that these pixels do not bias our measurements. 

In the estimation of the radial variation of mean SFR within galaxies, we performed a weighted mean of SFRs of the pixels within each radial annulus of each galaxy, out to r=1.5 Petrosian radii ($1.5R_p$) in the $r'$ band. Here, in calculating the total SFR, we also limit the spatial extent within which the SFR in the pixels is counted. The total SFR for each galaxy is therefore calculated from the sum of SFRs in all pixels in the galaxy, summed within consecutive radial annuli from the center of the galaxy up to $1.5R_p$. We find that estimating the total SFR within $1.5 R_p$ captures the vast majority of high signal-to-noise pixels that actually belong to the galaxy while not including the contribution of very low signal-to-noise pixels in the outskirts of the galaxy. While low surface brightness, low signal-to-noise pixels in the galaxy outskirts should individually have only a very small contribution to the total SFR (since we are weighting the SFR in each pixel by the square of the signal-to-noise), galaxies can have a substantially large areas of these low signal-to-noise pixels which extend well beyond the disk scale length. The combined contribution of these pixels could therefore bias the total SFR that is calculated. This appears to be less of a problem for typical disk galaxies than for bulge-dominated systems which can have a large area of these low-surface brightness pixels surrounding the galaxy. The second motivation of reducing the spatial extent is to remove a secondary contaminant: a smaller number of pixels which extend even beyond $2R_p$ of the galaxy and are clearly sky pixels whose fluxes are still fitted by the algorithm. Limiting the spatial extent to $1.5R_p$ within which the SFR in the pixels is counted, as was done in the radial analysis, therefore provides a more accurate estimate of the total SFR in the galaxy. 

\begin{figure}
\centering
\includegraphics[width=.63\textwidth,height=0.35\textwidth]{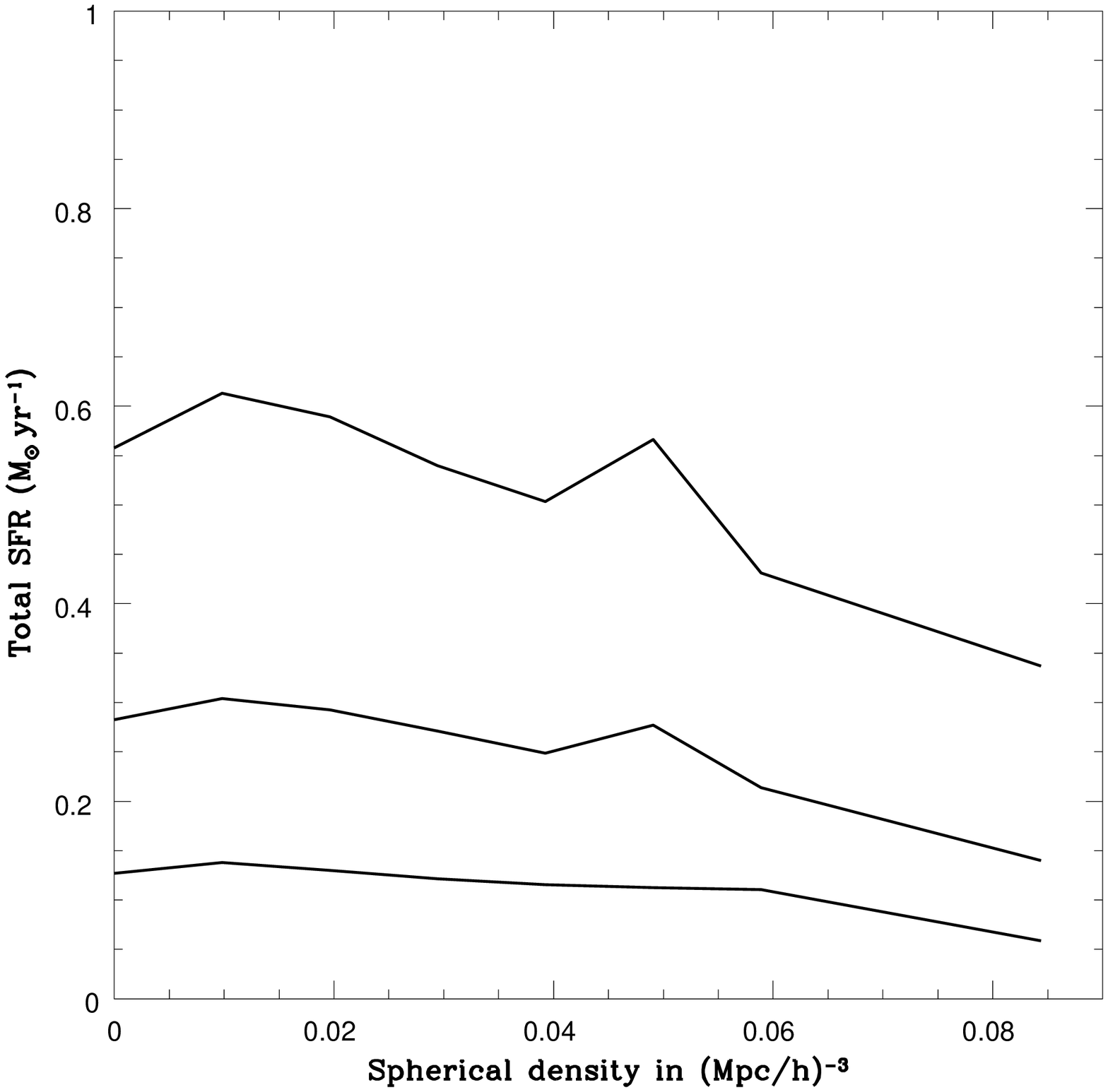}
\hfill
\includegraphics[width=.63\textwidth,height=0.35\textwidth]{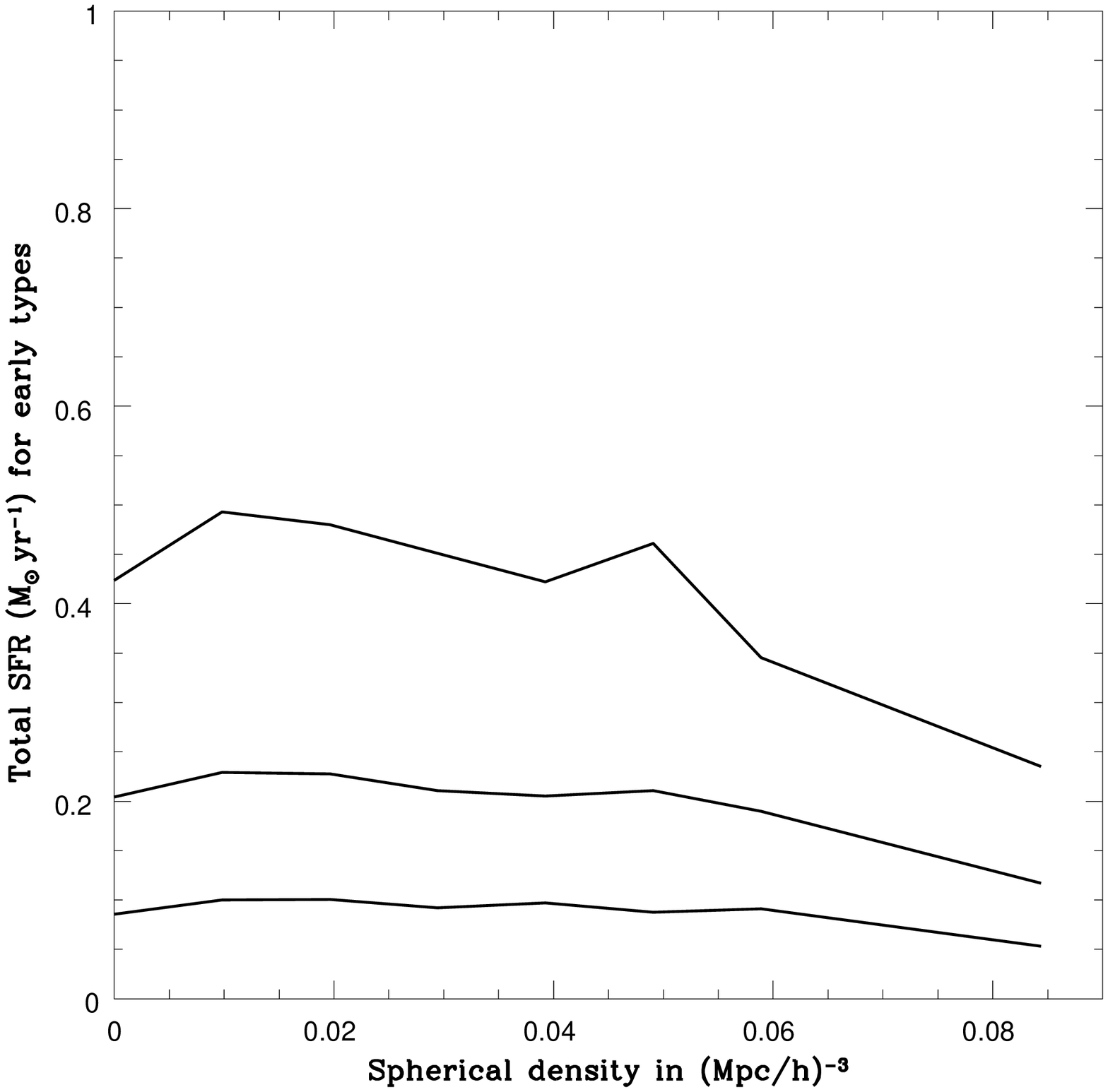}
\hfill
\includegraphics[width=.63\textwidth,height=0.35\textwidth]{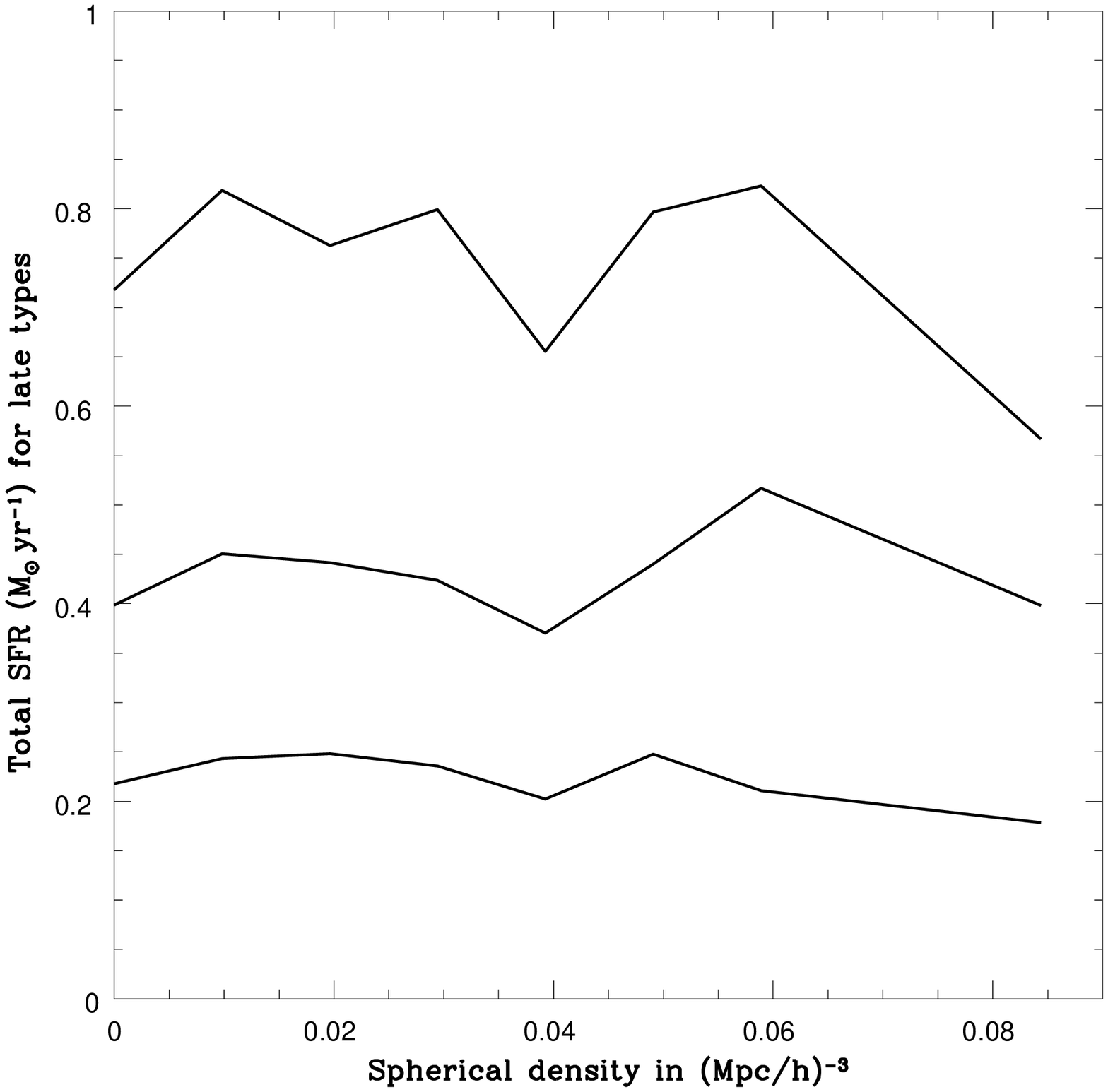}
\caption{Galaxy SFR in $M_{\odot} \, yr^{-1}$ as a function of density, where the SFR is calculated from a weighted sum over pixels in each galaxy. Top panel: for all types. Middle panel: for early-type galaxies. Bottom panel: for late-type galaxies. Unlike the analysis in Paper I, the spatial extent is constrained to $1.5\,R_p$ within which the SFRs in the pixels of galaxies are counted. The lines indicate the 75th, median and 25th percentiles of the SFR distribution respectively. Late-type galaxies now dominate the tail of the total SFR distribution at all densities. The trends with environment are identical to those found in Paper I, Figure 8 since only a higher fraction of low signal-to-noise and sky pixels are removed here. 
The decrease at higher densities is most noticeable in the most strongly star-forming galaxies, those in the 75th percentile of the total SFR distribution. 
\label{fig:totalsfr}}
\end{figure}

Figure~\ref{fig:totalsfr} shows the variation of the total SFR distribution
with local density for the full sample of galaxies (all types) and for the early-type and late-type subsamples within this. The three lines in each plot correspond to the 25th, median
and 75th percentiles of the SFR distribution. Removing a larger component of the low signal-to-noise pixels results in a more marked separation between the SFR distributions of early and late-type galaxies at all densities. Late-type galaxies now dominate the tail of the total SFR distribution -- they are seen to be more highly star-forming than early-types at all densities. The trends with environment are the same as the ones found in Paper I, Figure 9. It can be seen that in all three samples, the total SFR in galaxies is still correlated with the local environment as was found by \citet{Lew:02} and \citet{Gom:03}. As stated in Paper I, the fluctuations at low densities are characteristic of the size of the systematic uncertainties in
these measurement. In all three samples, the SFR decreases with increasing density,
with the greatest effect in the highest density environments: $>0.05\,$($h^{-1}\,Mpc$)$^{-3}$ for early-types and for the full sample and $>0.055\,$($h^{-1}\,Mpc$)$^{-3}$ for late-types (these densities correspond to the outskirts of rich clusters). In each of the three samples, the effect is most noticeable in the most strongly star-forming galaxies, i.e.\ those in the 75th percentile of the SFR distribution. The SFR distributions in all three samples are skewed towards higher SFRs at low densities. As density increases, the skewness of the SFR distributions in the early and late-type galaxies decreases. It is also worth noting that even in the highest density environments, the SFR of these high-SF late-type galaxies is higher than for the early-types.

 Limiting the spatial extent to $1.5R_p$ over which the total SFR in galaxies is estimated, as well as performing a weighted sum over the pixels, thus removes a source of contamination that was more significant that previously thought. This results in a clearer separation between early and late-type galaxies in terms of the total SFR distribution. The trends between SFR and environment, however, remain largely intact. 


\section{Radial and environmental trends in SFR}
\label{sec:results}

\subsection{The Density-Morphology Relation}
\label{subsec:dens_morph}

\begin{figure}
\centerline{\rotatebox{0}{
\includegraphics[width=.55\textwidth,height=0.55\textwidth]{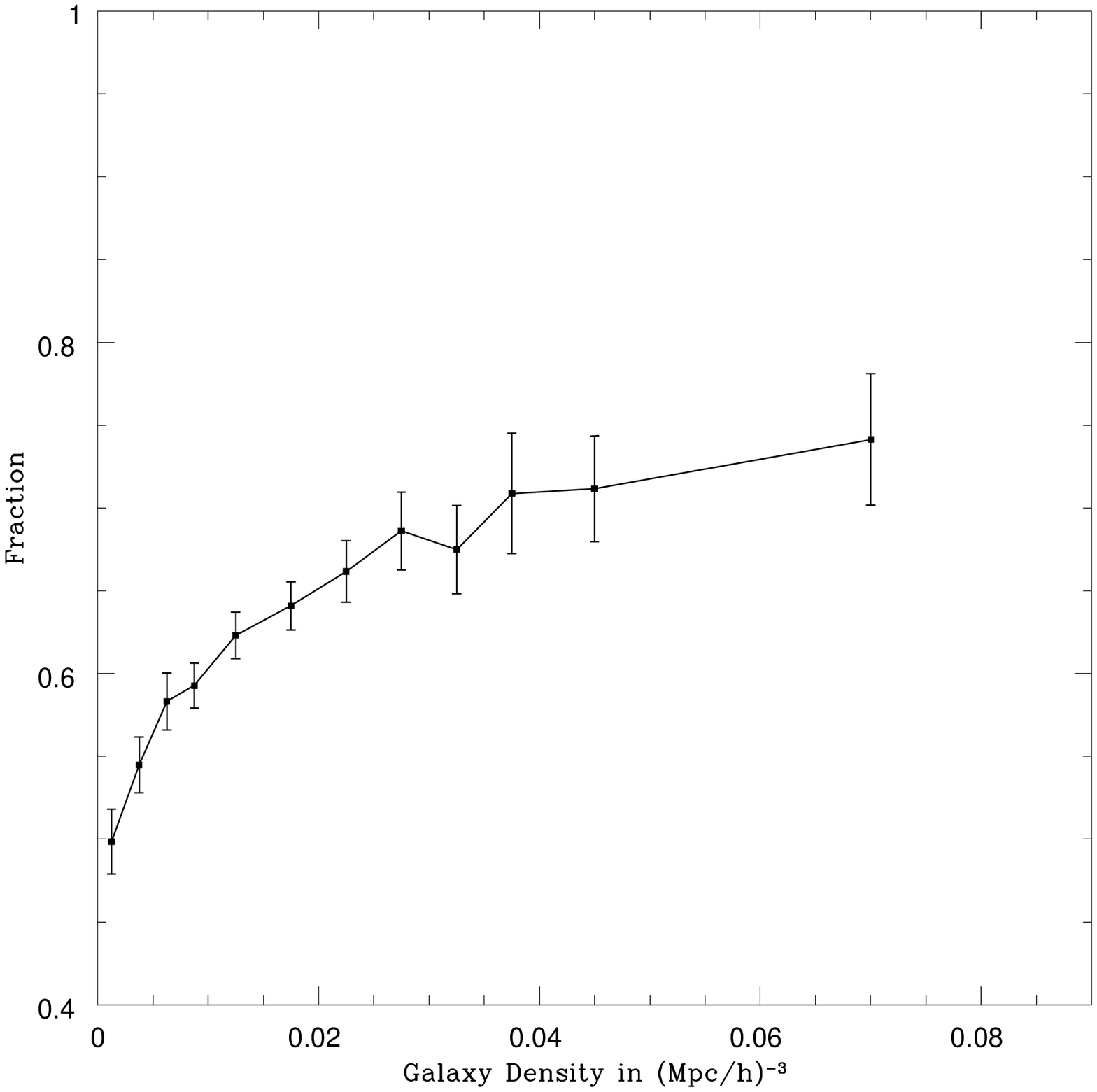}
}
}
\vspace{0.3cm}
\centerline{\rotatebox{0}{
\includegraphics[width=.55\textwidth,height=0.55\textwidth]{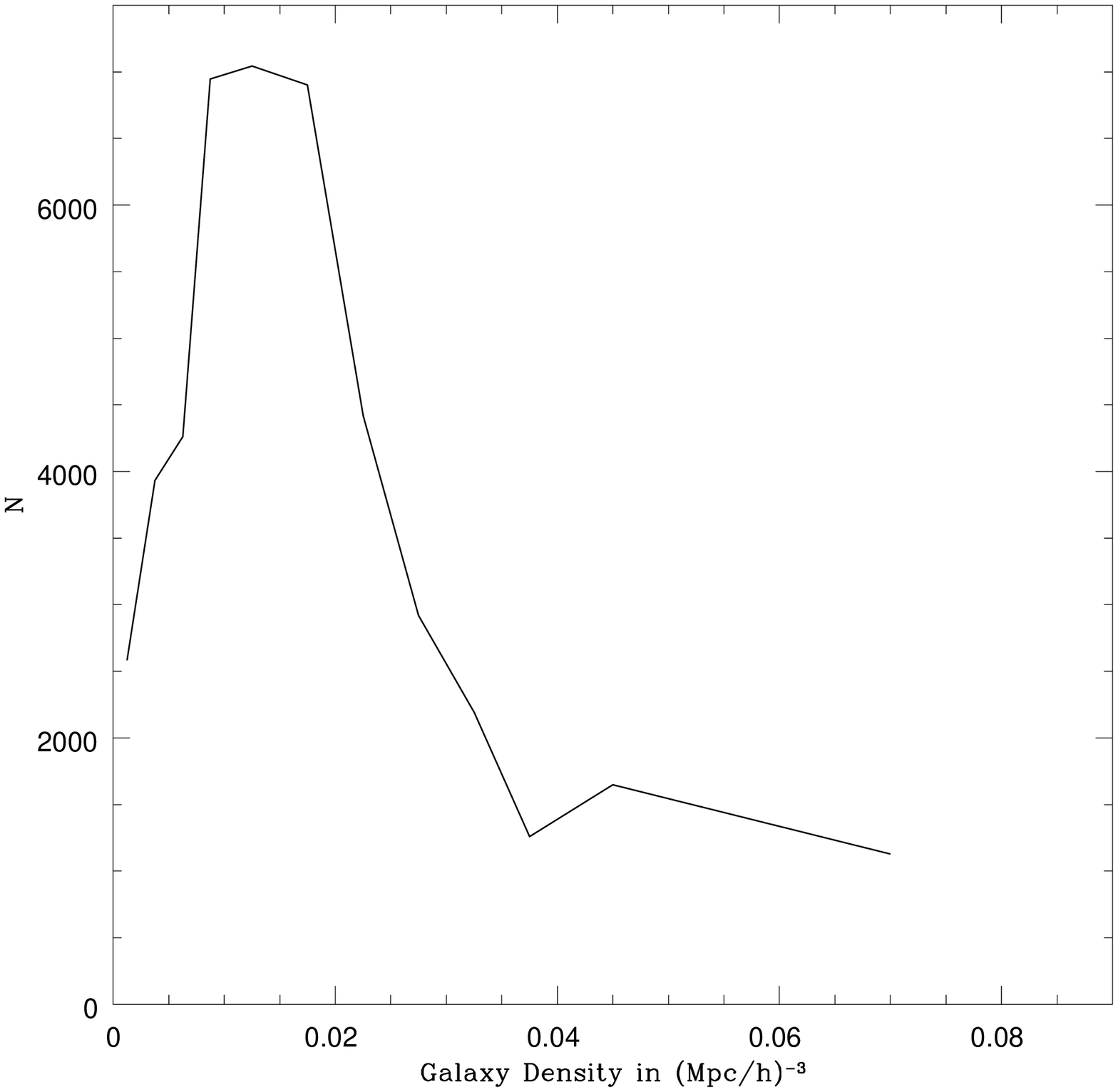}
}
}
\caption{Top panel: The density-morphology relation, showing the fraction of early-type galaxies (with $C_{in}<0.4$). Error bars are Poisson. Bottom panel: Number of galaxies in each interval of local galaxy density. The local galaxy density is sampled more finely in these bins than in the subsequent analysis, where only three intervals of local density are used. \label{fig:DENSITY-MORPH}}
\end{figure}

Figure~\ref{fig:DENSITY-MORPH} illustrates the density-morphology
relation in this sample of SDSS galaxies. The proportion of early-type
galaxies, with $C_{in}<0.4$, is shown to be increasing as a function of
local galaxy density. In the low-density environments
around $50$ percent of galaxies are early-types, while in the highest
density environments this increases to around $75$ percent. The bottom
panel gives the total number of galaxies found in each density
interval. These plots sample the local galaxy density more
finely that in the analysis of Paper I, or in subsequent analyses herein
where only three intervals of local density are used. We use the results of
the density-morphology relation below in \S\,\ref{subsec:density_morph}
when assessing whether the observed suppression in SF can be recovered simply
by mimicking this effect. Before we can do this, though, we need to establish
the radial variation of SFR for each morphological type as a function of
environment.

\subsection{Radial Variation of SFR as a Function of Environment
\label{subsec:radial_types}}

Here we examine the radial distribution of SFR in both early-type and
late-type galaxies as a function of their environment for galaxies
(a)~spanning the entire range of (total) SFR and (b)~within the highly
star-forming population only.



\begin{figure}
\centering
\includegraphics[width=.63\textwidth,height=0.4\textwidth]{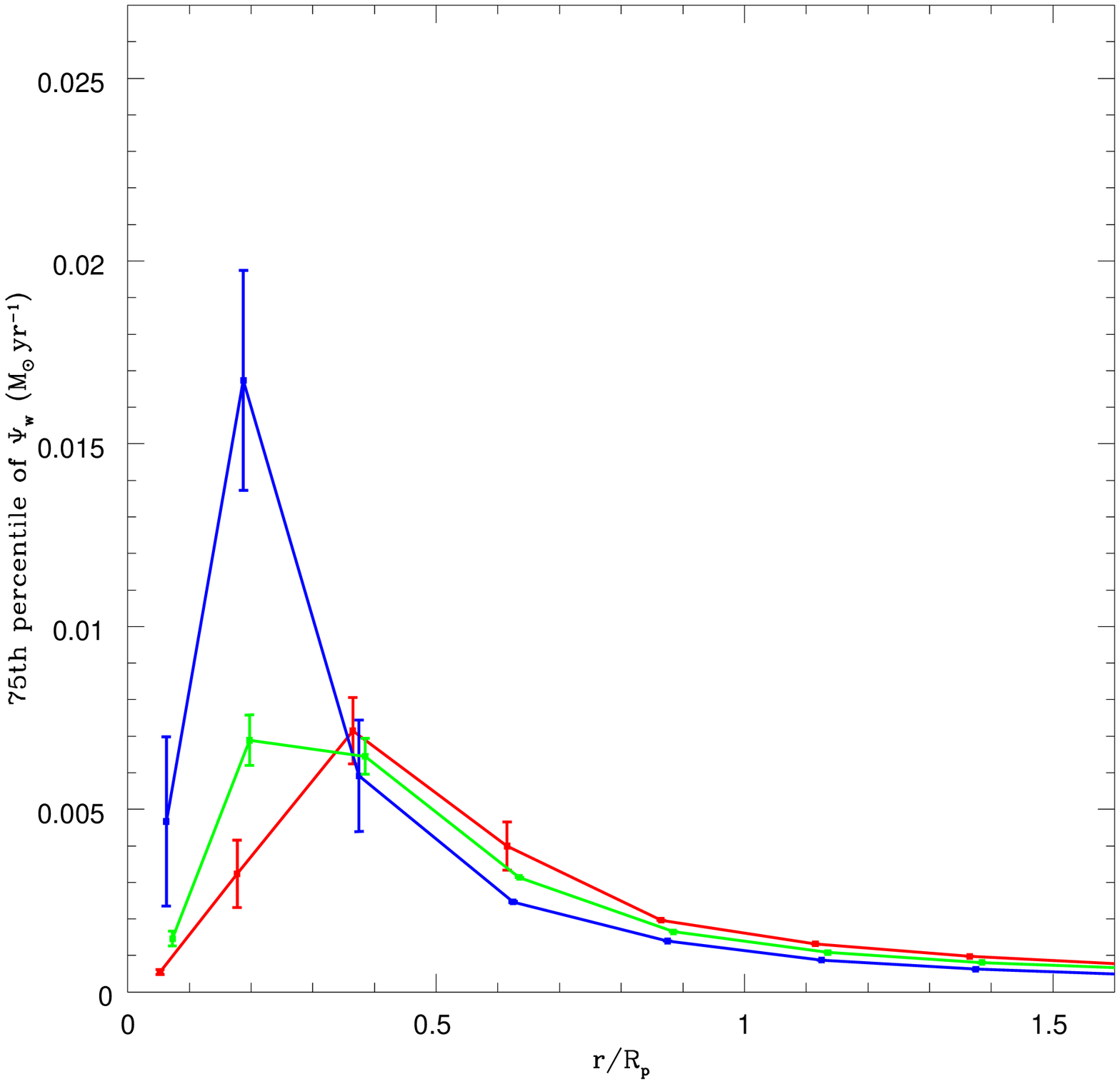}
\hfill
\includegraphics[width=.63\textwidth,height=0.4\textwidth]{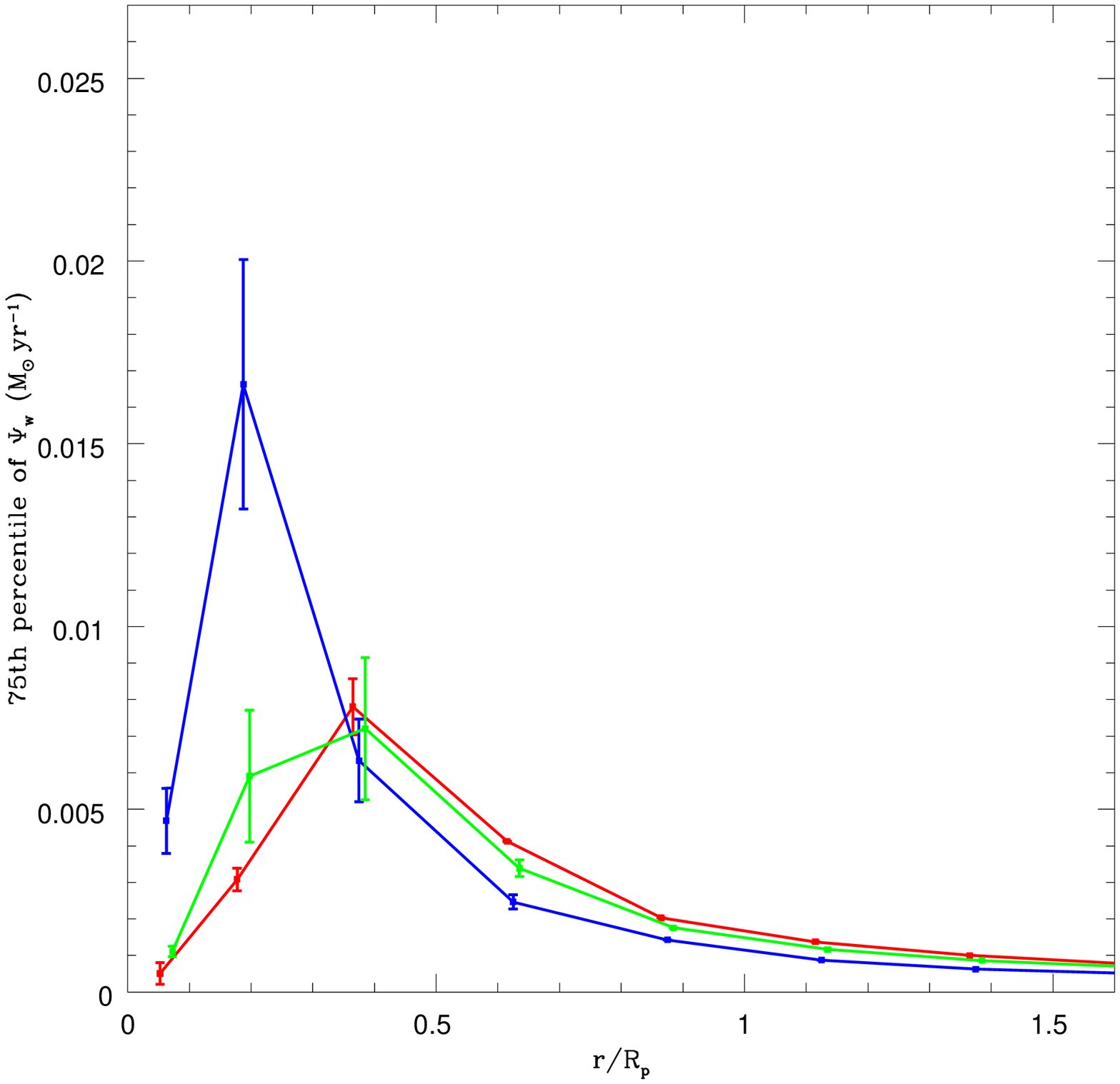}
\hfill
\includegraphics[width=.63\textwidth,height=0.4\textwidth]{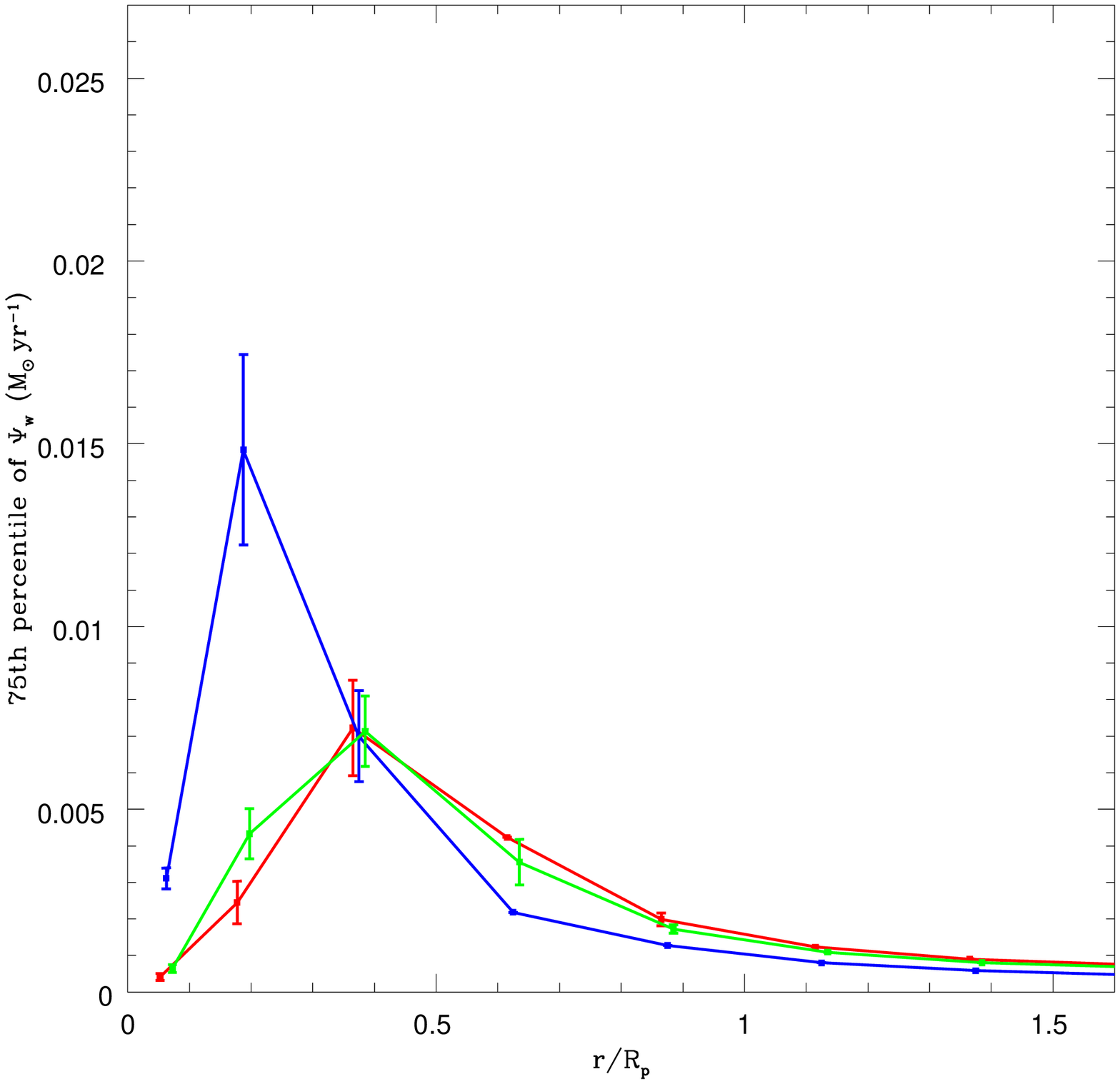}
\caption{Top panel: 75th percentiles of the distribution of weighted
mean SFRs $\Psi_{w}$ ($M_{\odot} \,yr^{-1} $) within successive radial
annuli for early-type galaxies (red), late-type galaxies (blue), and
all galaxies (green) which have local galaxy densities $\rho$ in the
range $0.0 < \rho \le 0.01 \,$(Mpc/h)$^{-3}$. Middle panel: $0.01 <
\rho \le 0.04 \,$(Mpc/h)$^{-3}$. Bottom panel: $0.04 < \rho \le 0.09
\,$(Mpc/h)$^{-3}$. The three density intervals are the same as those used
in Paper~I, Figure~9.
\label{fig:radialplots_densities}}
\end{figure}

We calculate the weighted mean SFR $\psi_w$ for each radial annulus
in every galaxy
(see Paper I for details). For a sample of galaxies this gives
a distribution of $\psi_w$ for each annulus.
Figure~\ref{fig:radialplots_densities} shows the 75th
percentile of $\psi_w$ for the early-types, late-types and for the full
sample, within three intervals of galaxy density (corresponding to the three
panels in the Figure). As shown in Paper I the effect of environment is most
pronounced for the 75th percentile of
$\psi_w$, rather than the median of $\psi_w$. In determining the radial
variation of SFR and its dependence on environment we focus on the 75th
percentile of $\psi_w$ in this study.

{\em Both early and late-type galaxies have distinct radial SFR profiles.}
This answers the first of the questions we posed above. The
peak of $\psi_w$ in early-types is significantly lower than in late-type
galaxies. The SF in the center is very low, reflecting the
fact that in early-type systems, which are typically bulge-dominated,
the stellar populations in the center of the bulge are old and there
is consequently little ongoing SF. In contrast, the stellar populations with
higher SFR are further out in the bulge and the mean SF in this region is
also better sampled.
The early-type galaxies show a more extended distribution of SF which peaks
further out in radius than in late-type galaxies. This reflects the fact
that late-type galaxies have a smaller bulge compared to early-types.

In the late-types most of the SF takes places in the inner part of the
disk, closer to the center. There is thus a very sharp increase in SFR as
we go from the center ($r \le 0.125\,R_p$) to the inner part of the disk
($r \approx 0.25\,R_p$) followed by a rapid decline in the SFR throughout
the disk to the outskirts. Disk galaxies have, on average, a much higher
SFR up to $r \approx 0.25\,R_p$ (by as much as $0.014\,M_{\odot}\,$yr$^{-1}$
in the lowest density environments and $0.012\,M_{\odot}\,$yr$^{-1}$
in the highest density environments) than do the early-types. This reflects
the younger star-forming stellar populations in the disks of these galaxies.

As described in Paper I, the effect of galaxy density on the full sample (all types) is to cause a supression of SF in the centers of galaxies ($r<0.25R_p$) while no effect is observed for $r>0.25R_p$. Figure~\ref{fig:radialplots_densities} also shows that the effect of the environment on the full sample of galaxies (early and late types together, green curve) is to push the peak of the SFR profile to larger radii. However, Figure~\ref{fig:radialplots_densities} also shows that when galaxies in the full sample are divided by type, the SFR radial profiles of early and late-type galaxies do not change significantly across the different environmental ranges. 

A straightforward explanation of the trend in the full sample in terms of the density-morphology relation presents itself, namely that at low densities there is almost an equal fraction of early and late-type galaxies, while at higher densities the
profile is progressively more dominated by the higher contribution from
early-types. It is important to note that while the shifting of
the peak mean SFR in the full sample of galaxies may indeed be explained
simply by the increasing fraction of early-type galaxies in higher density
environments, there are other aspects of the density dependence that
are not.

\subsection{The Significance of Highly Star Forming Galaxies
\label{subsec:highsf_subpop}}

In Paper I it was found that the most significant decrease in the
total SFR with increasing local galaxy density takes place for galaxies
in the top quartile of the total SFR distribution. Here we explore the
radial SFR profiles for the galaxies that are in the top and
second quartile of the total SFR distribution for each galaxy type, to identify whether the
trends with environment are driven only by the most active star-forming
galaxies. In \S\,\ref{subsec:luminosity}, we also examine the proposal by \citet{Par:07} that the trends are instead driven primarily by the most luminous galaxies. 

We refer to the top quartile of the total SFR distribution, for both
morphological types, as the ``highest SF galaxies,'' and the second quartile
as the ``the next highest SF galaxies''.
We examine the distribution of $\psi_w$ for these subpopulations for early
and late-types separately, determining the quartiles of $\psi_w$ of
these distributions as above in the same intervals of local galaxy density.
The results, as before, are most prominent in the 75th percentile of $\psi_w$
(compared to the median or 25th percentile) and radial trends for only this
quartile are shown.

\begin{figure}
\centerline{\rotatebox{0}{
\includegraphics[width=.55\textwidth,height=0.55\textwidth]{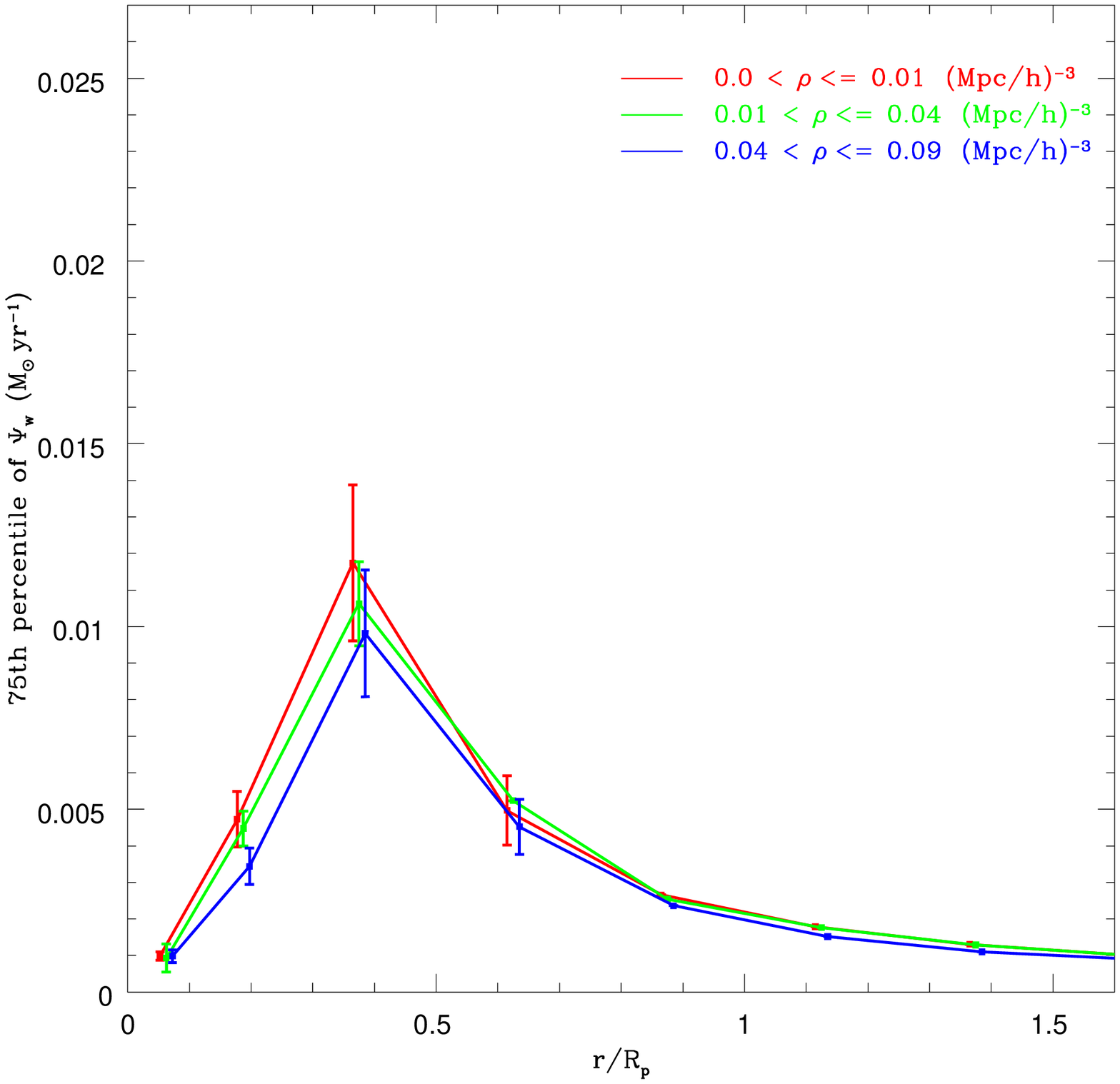}
\includegraphics[width=.55\textwidth,height=0.55\textwidth]{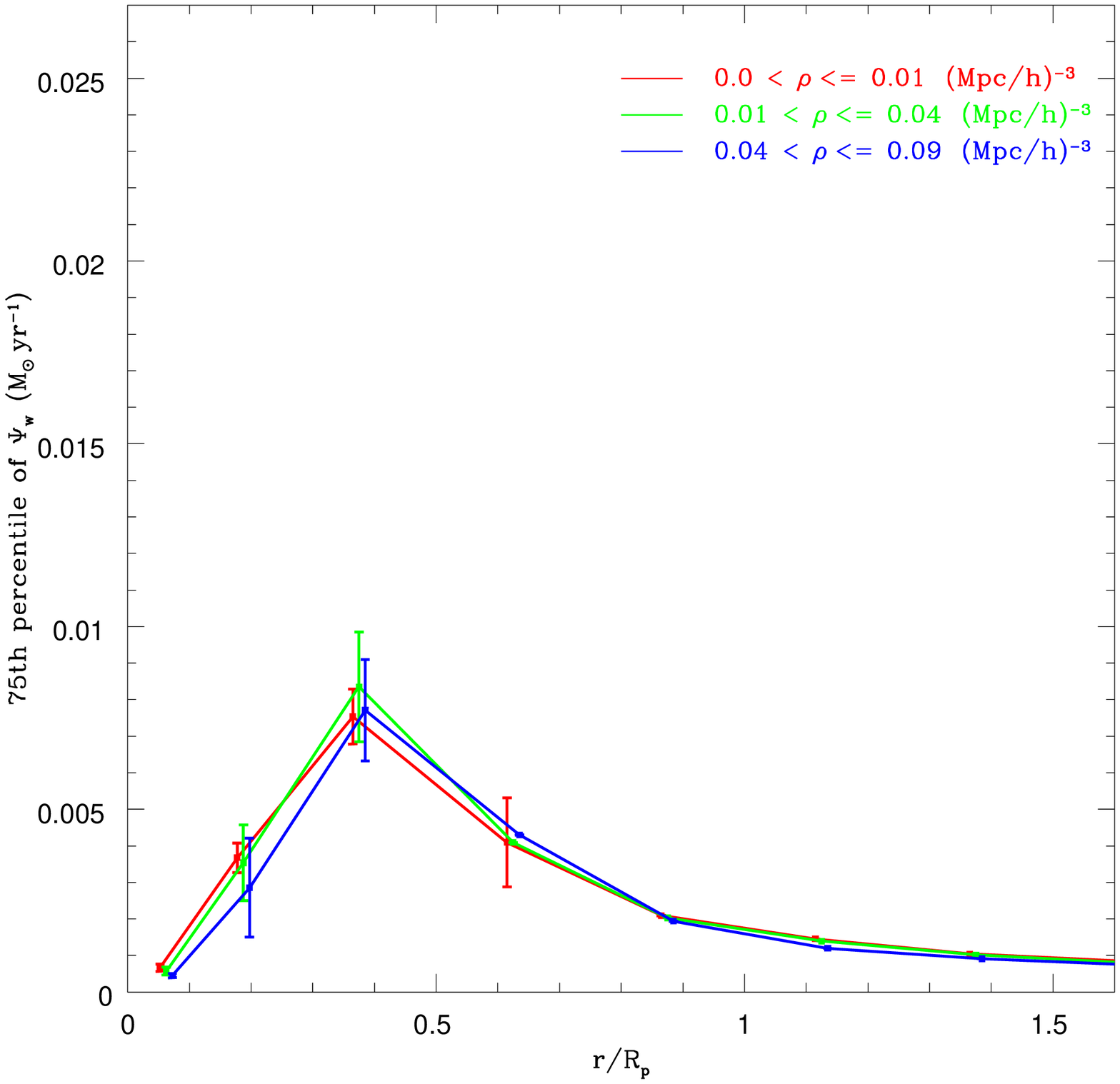}
}
}
\vspace{0.4cm}
\centerline{\rotatebox{0}{
\includegraphics[width=.55\textwidth,height=0.55\textwidth]{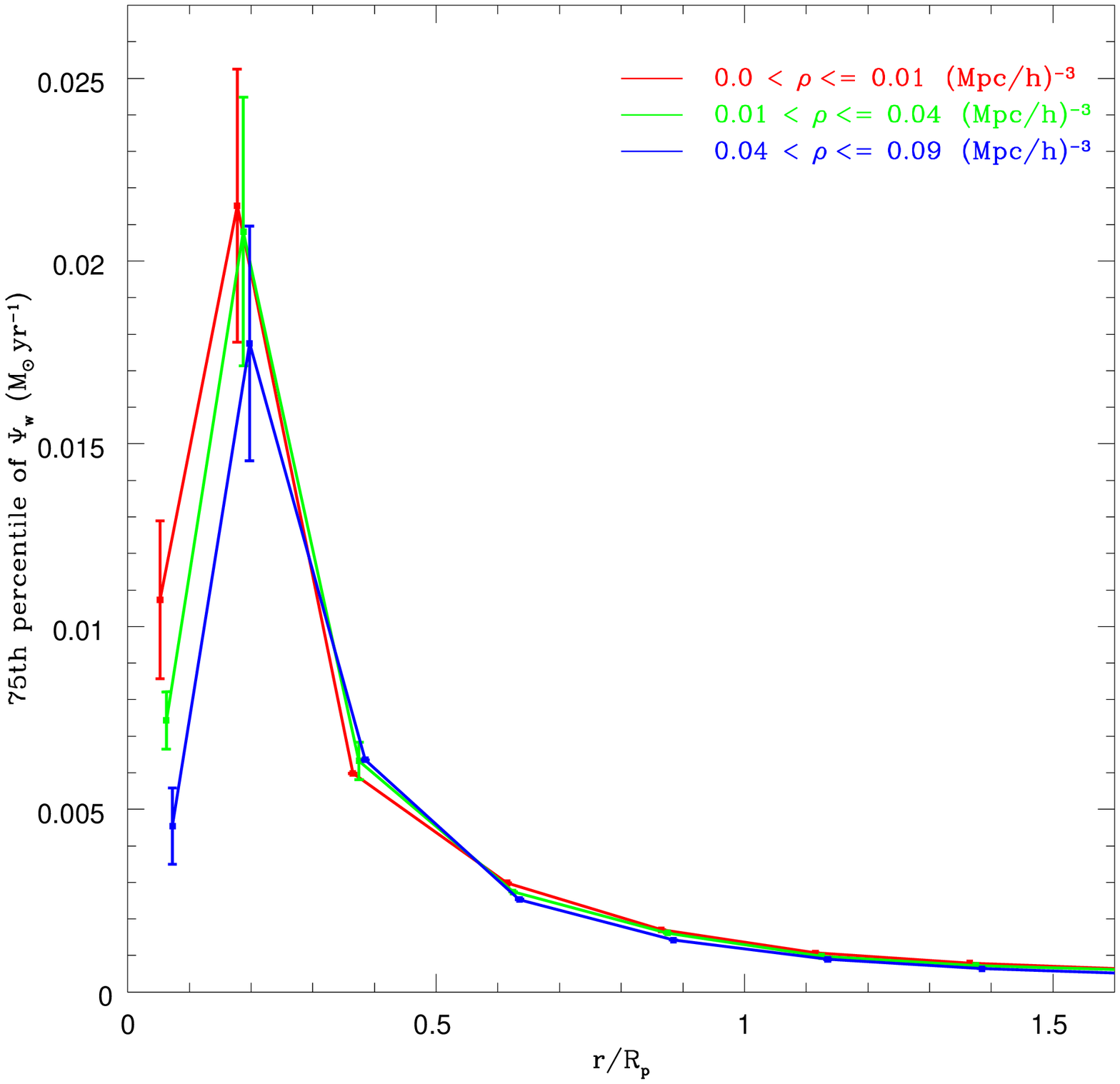}
\includegraphics[width=.55\textwidth,height=0.55\textwidth]{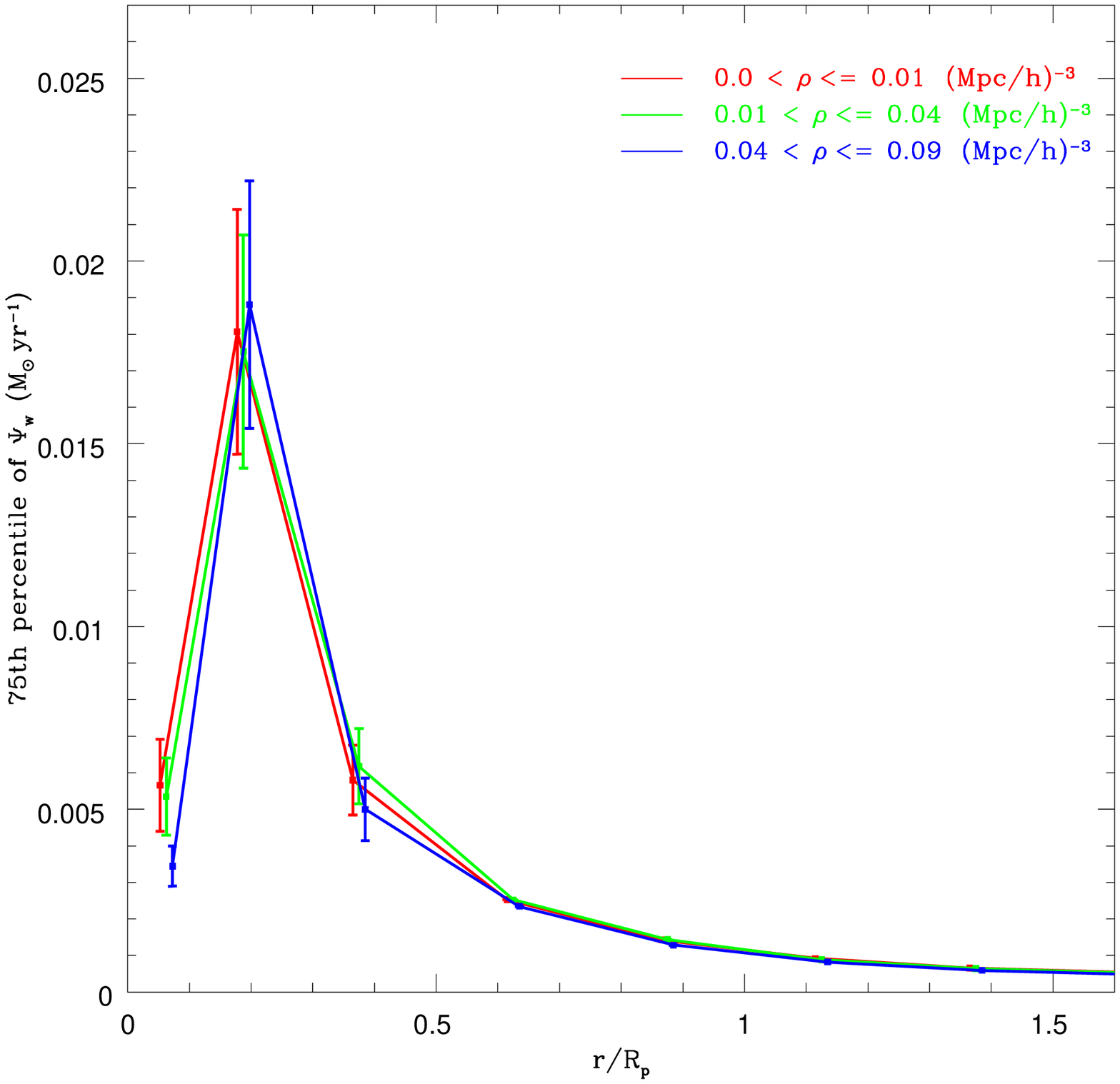}
}
}
\caption{75th percentiles of the distribution of weighted mean SFRs
$\Psi_{w}$ ($M_{\odot} \,yr^{-1} $) within successive radial annuli as a
function of the local galaxy density $\rho$ for the highly SF populations
of either galaxy type. The intervals of local galaxy density $\rho$
considered are $0.0 < \rho \le 0.01 \,$(Mpc/h)$^{-3}$ (red), $0.01 <
\rho \le 0.04 \,$(Mpc/h)$^{-3}$ (green), $0.04-0.09 \,$(Mpc/h)$^{-3}$
(blue). Top left panel: For the early-type galaxies in the highest quartile
of the total galaxy SFR (`the highest SF galaxies'). Top right panel:
For early-type galaxies in the second highest quartile of the total
galaxy SFR (`the next highest SF galaxies'). Bottom left panel: For the
highest SF late-type galaxies. Bottom right panel: For the next highest
SF late-type galaxies.
\label{fig:radialplots_highsf}}
\end{figure}

\begin{figure}
\centerline{\rotatebox{0}{
\includegraphics[width=.55\textwidth,height=0.55\textwidth]{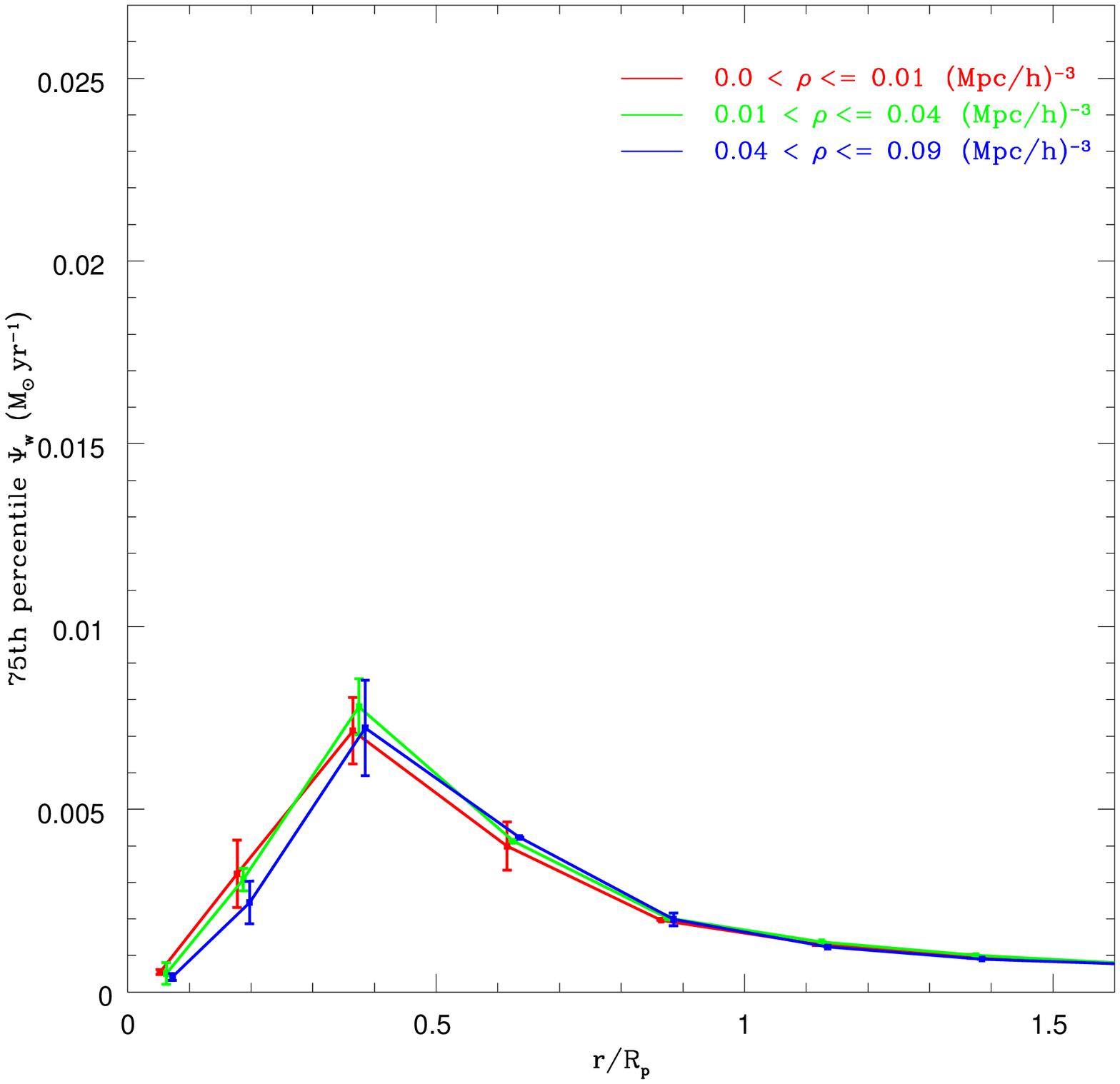}
\hspace{1.0cm}
\includegraphics[width=.55\textwidth,height=0.55\textwidth]{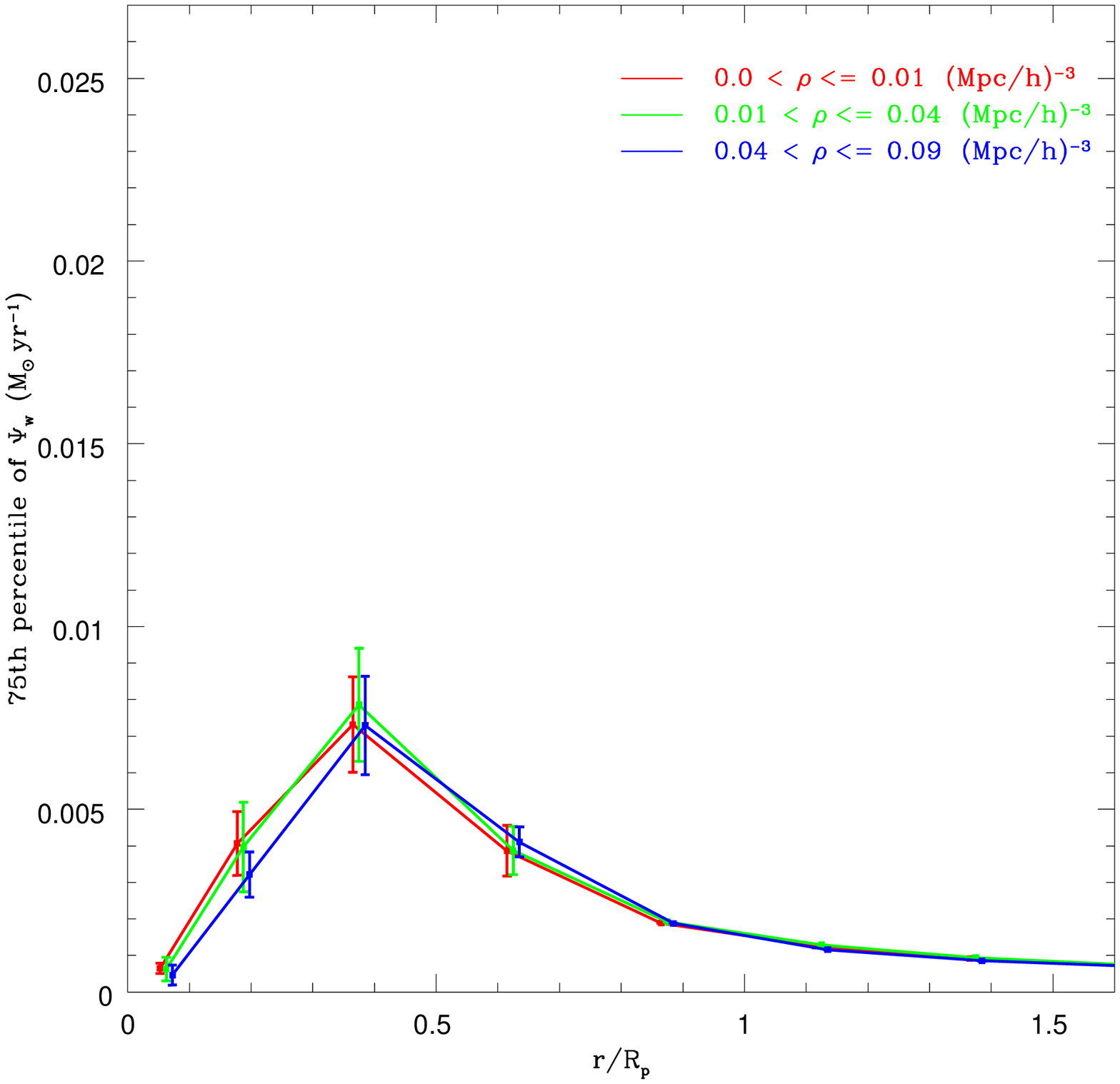}
}
}
\vspace{0.4cm}
\centerline{\rotatebox{0}{
\includegraphics[width=.55\textwidth,height=0.55\textwidth]{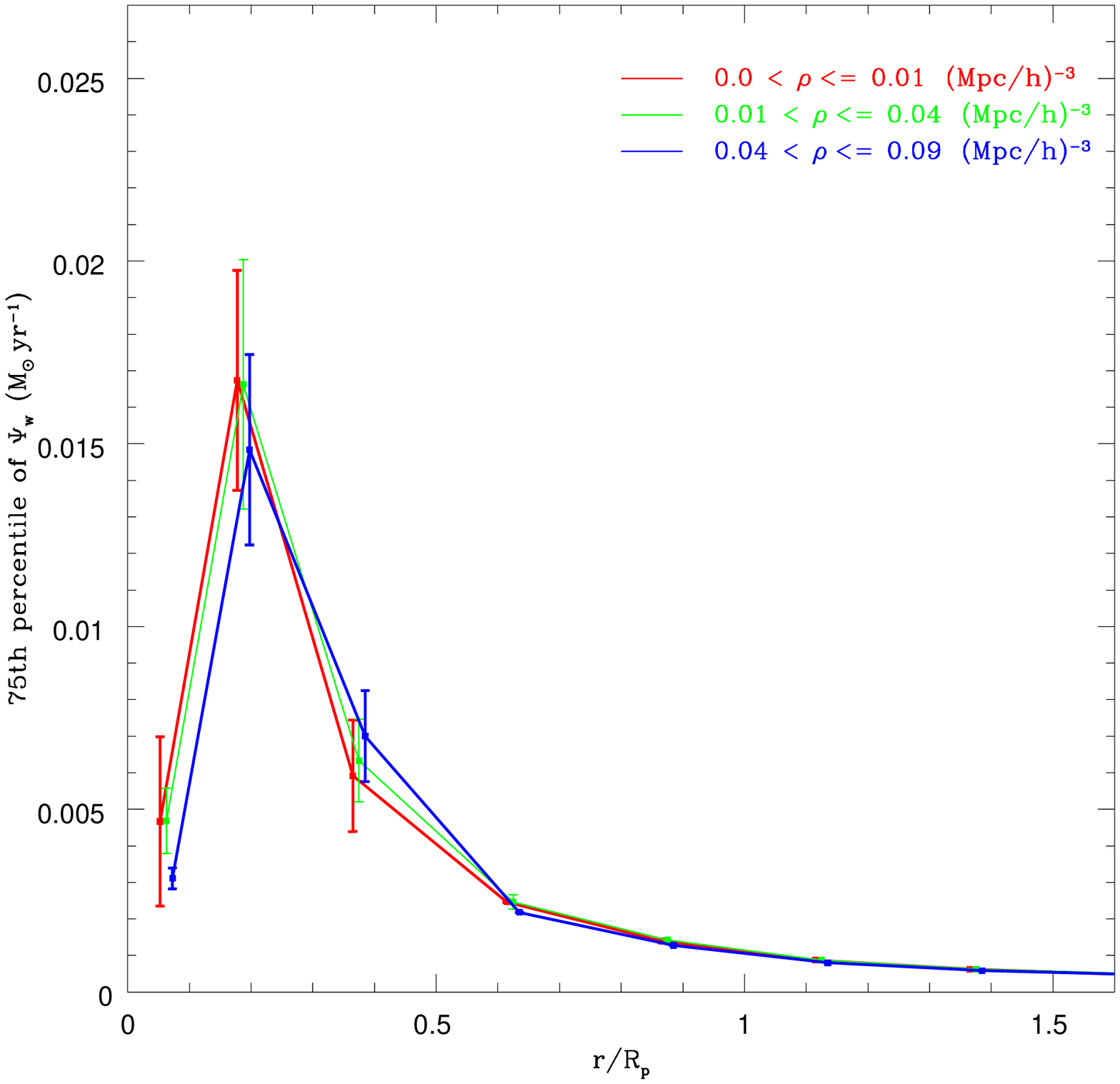}
\hspace{1.0cm}
\includegraphics[width=.55\textwidth,height=0.55\textwidth]{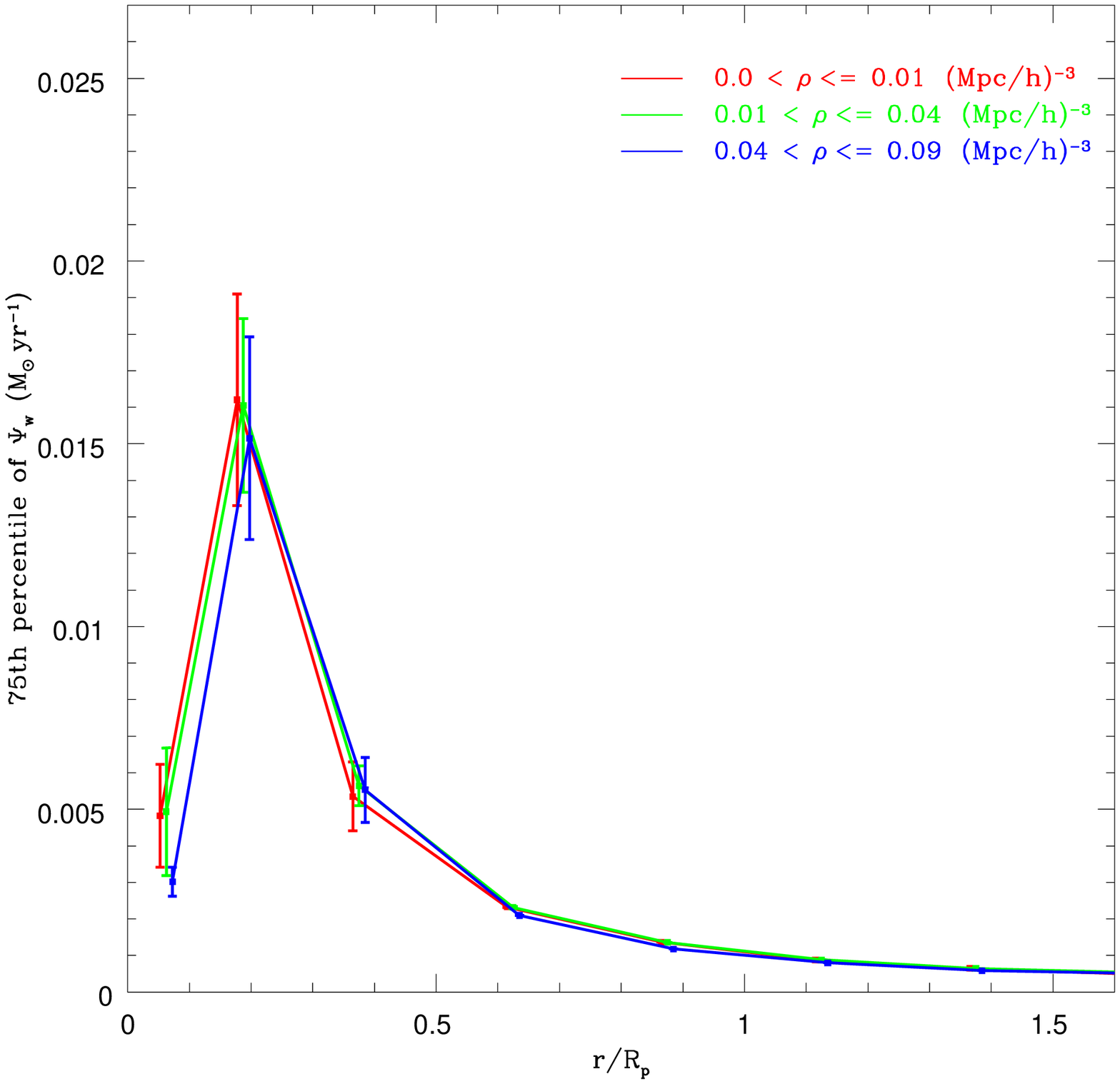}
}
} 
\caption{75th percentiles of the distribution of weighted mean SFRs
$\Psi_{w}$ ($M_{\odot} \,yr^{-1} $) within successive radial annuli as
a function of the local galaxy density $\rho$ for the full sample of
galaxies of either type. No cuts in total SFR are made here. Top left
panel: For all early-type galaxies chosen according to the inverse concentration
index ($C_{in}<0.4$). Top right panel: For all early-type galaxies
chosen according to the Sersic index ($n>2$). Bottom left panel:
late-type galaxies chosen using $C_{in}>0.4$. Bottom right panel:
late-type galaxies chosen using $n<2$.
\label{fig:radialplots}}
\end{figure}

The results of this analysis are presented in
Figure~\ref{fig:radialplots_highsf}, with the full sample spanning all SFRs
shown for comparison in Figure~\ref{fig:radialplots}.
Figure~\ref{fig:radialplots_highsf} shows
the 75th percentile of $\psi_w$ for the highest and next highest SF
galaxies for both galaxy types. In the early-type highest-SF galaxies
there is a relatively small suppression ($1.5\sigma$) in $\psi_w$ between
the least and most dense environments but only in the region $0.125 <
r/R_p \le 0.25$. We do not see a significant effect of environment on
the SFR in early-types in the center ($0.0 < r/R_p \le 0.25$) or in
the outskirts. In the next highest SF early-types, we do not detect
a similar suppression in the SF, suggesting that this is a phenomenon
that affects only the most active early-type systems. In the highest-SF
late-type galaxies, we detect a suppression of $3\sigma$ in the mean
SFR in the center ($r/R_p \le 0.125$), while there is no significant
suppression of SF at other radii. In the next highest SF late-types,
we find a slightly lower suppression in $\psi_w$ of $2\sigma$ in the SFR
in the center ($r/R_p \le 0.125$), and again, no significant effect
of the environment at other radii. The cores of late-type galaxies
thus have their SF reduced in more dense environments, while the SF
in the remainder of the disk and outskirts is largely unaffected by a
changing environment. Further, unlike the early-type systems, this
affects a larger proportion of star-forming late-types. All late-types
in the top two quartiles of the total SFR distribution have their central SFR suppressed
in the highest densities. In contrast, in the early-types, we detect a
suppression only in galaxies in the highest quartile of the total SFR
distribution.

 A Kolmogorov-Smirnov (KS) test on the distribution of $\Psi_w$ of the two innermost annuli in both the lowest density interval ($0.0 < \rho \le 0.01 \,$($h^{-1}\,Mpc$)$^{-3}$) and the highest density interval ($0.04 < \rho \le 0.09 \,$($h^{-1}\,Mpc$)$^{-3}$) rules out the hypothesis that the populations in the two density ranges are derived from the same underlying distribution at more than $99\%$ confidence level. This holds for the full sample of highly star-forming galaxies or for early and late-type galaxies considered separately. The observed suppression of SF in the central regions of these galaxies is therefore an effect of the environment rather than being due to stochastic fluctuations.

Figure~\ref{fig:radialplots} shows the results of performing a similar
analysis for the full sample of early and late-type galaxies, without
making any cuts in total SFR. Neither galaxy type shows any statistically
significant variation in the 75th percentile of $\psi_w$ with environment
at any radius when the full sample of galaxies of each type is considered.
{\em The suppression of SF in the highest density environments is therefore
driven by the most active, highly SF galaxies.} This answers the second of
the questions posed above.
It is also worth noting that, as found for the full sample in Paper I
when considering both morphological types together, the SFR in the outskirts
of either the early or late-type galaxies is not affected by a changing
environment. The right-hand panels of Figure~\ref{fig:radialplots} also
show the radial SFR profiles when the Sersic index is used as a proxy
for galaxy morphology. It is clear that these results hold independent of
the choice of inverse concentration index $C_{in}$ or the Sersic index as the
morphological proxy.

\subsection{Can the density-morphology relation alone explain the suppression
of star formation?
\label{subsec:density_morph}}

The results for the high SF populations of either galaxy type suggest
that the suppression of SF is not due solely to the density-morphology
relation. In order to determine definitively whether the suppression
in SF is a result of the density-morphology relation or if another,
perhaps evolutionary, mechanism is at work, we carry out a further
test. We have determined the radial SFR profiles for the early and late-types
in the lowest density environments, together with the fraction of early and
late-types as a function of density. So we can now ask how the radial SFR
profile compares for the full sample (all types) in the highest density
environment with a profile constructed by combining the low-density profiles
for each morphological type in the proportions appropriate to the high-density
environment.

\begin{figure}
\centerline{\rotatebox{0}{
\includegraphics[width=.55\textwidth,height=0.55\textwidth]{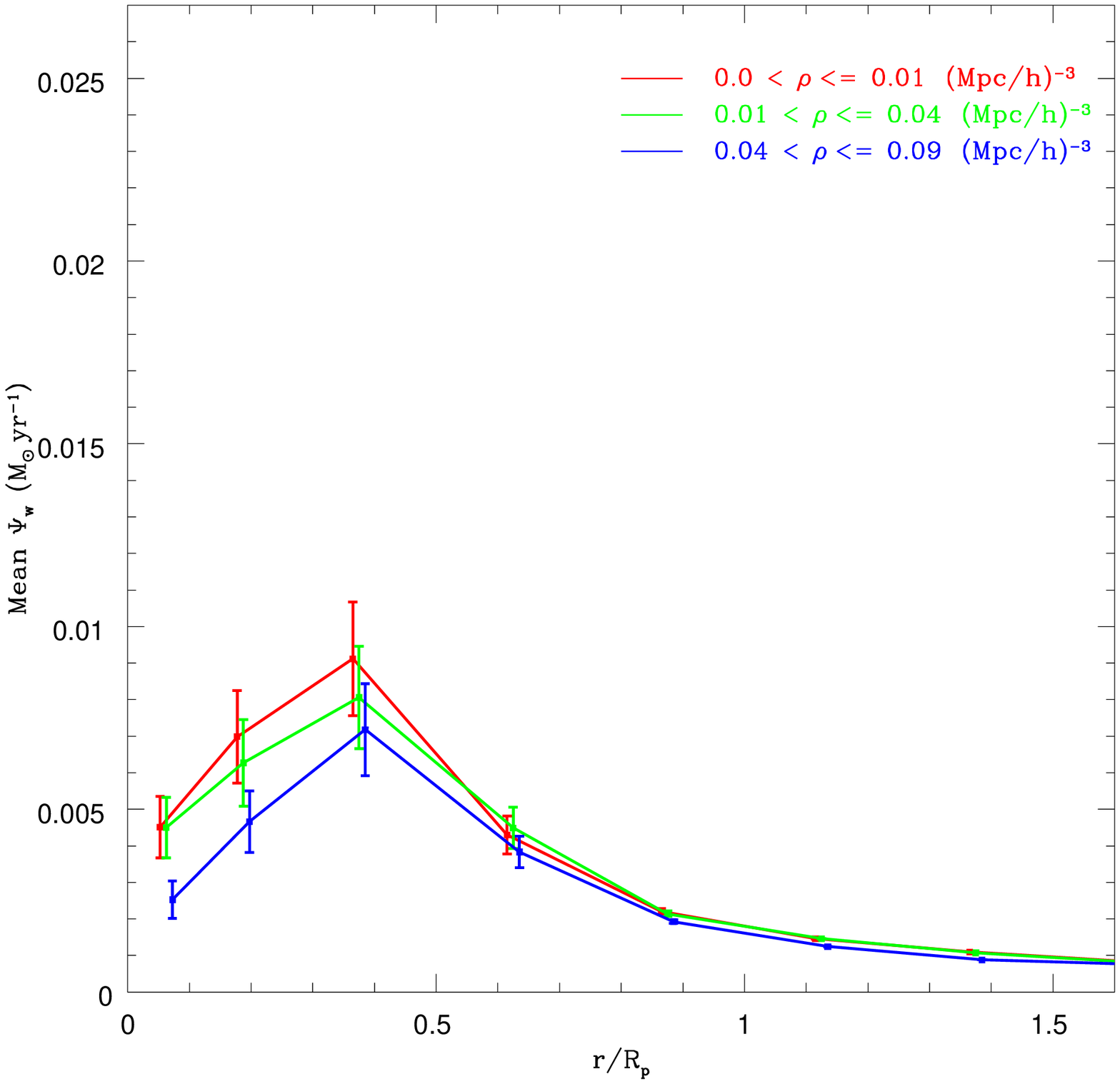}
\includegraphics[width=.55\textwidth,height=0.55\textwidth]{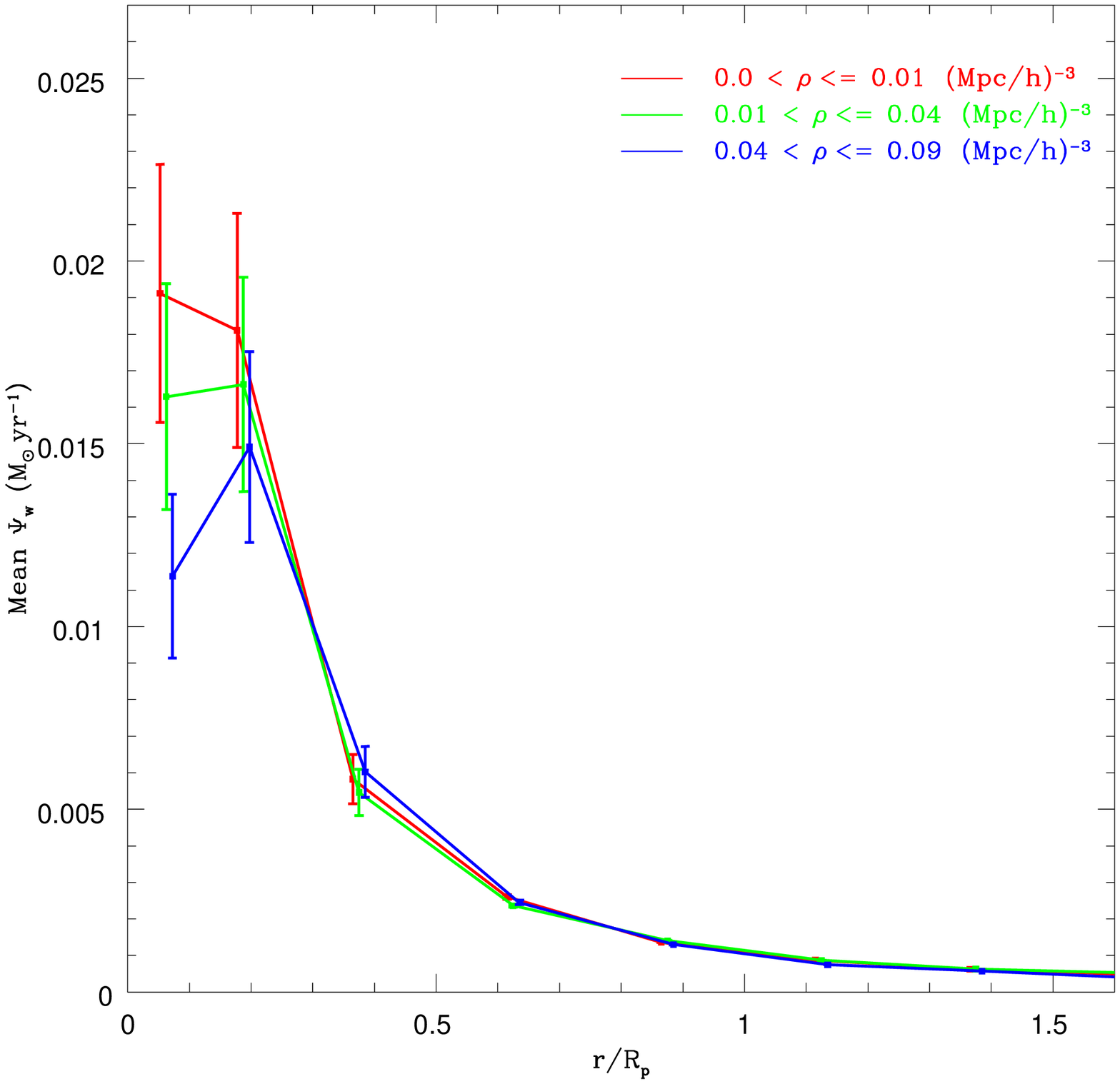}
}
}
\vspace{0.4cm}
\centerline{\rotatebox{0}{
\includegraphics[width=.55\textwidth,height=0.55\textwidth]{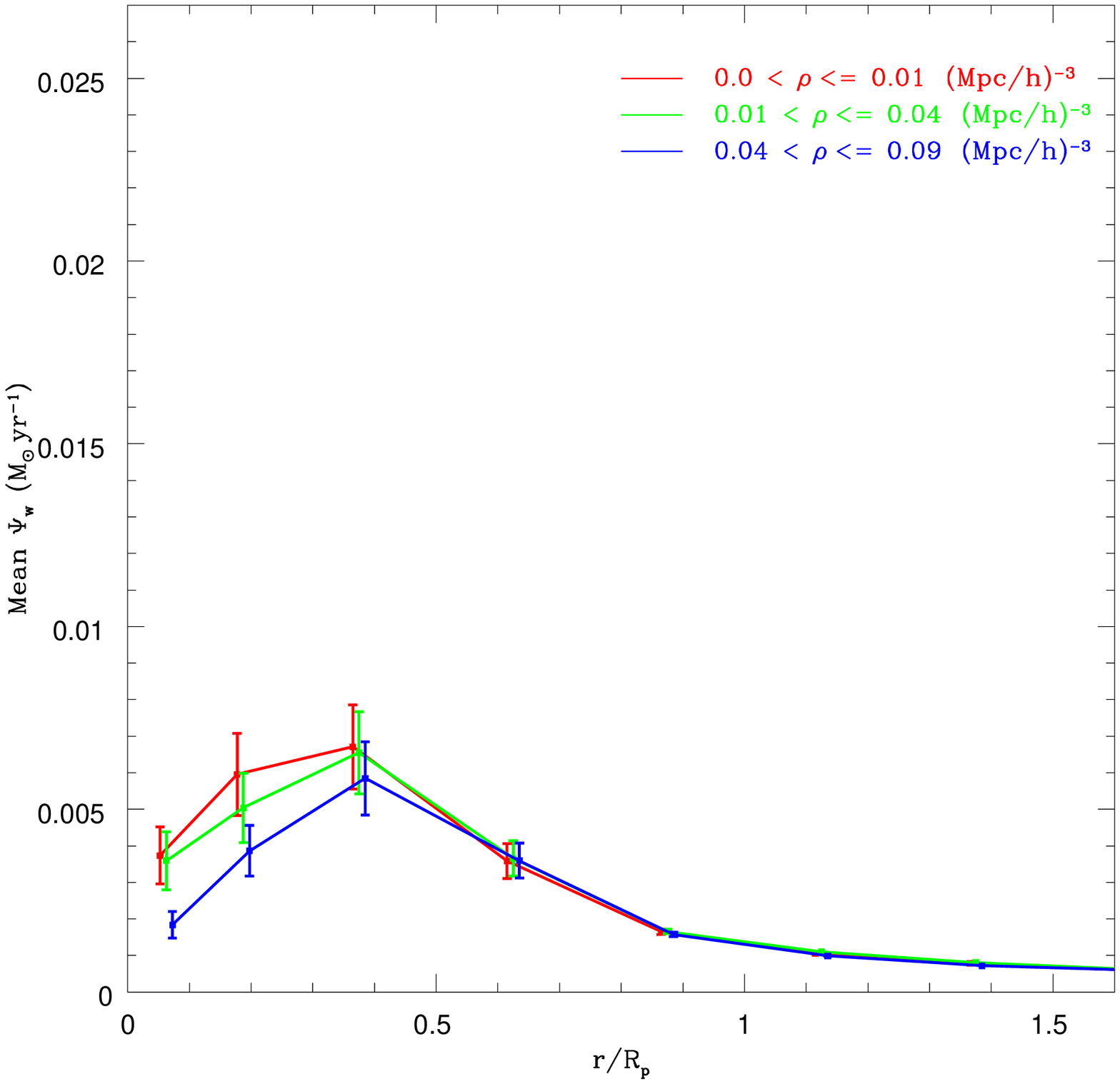}
\includegraphics[width=.55\textwidth,height=0.55\textwidth]{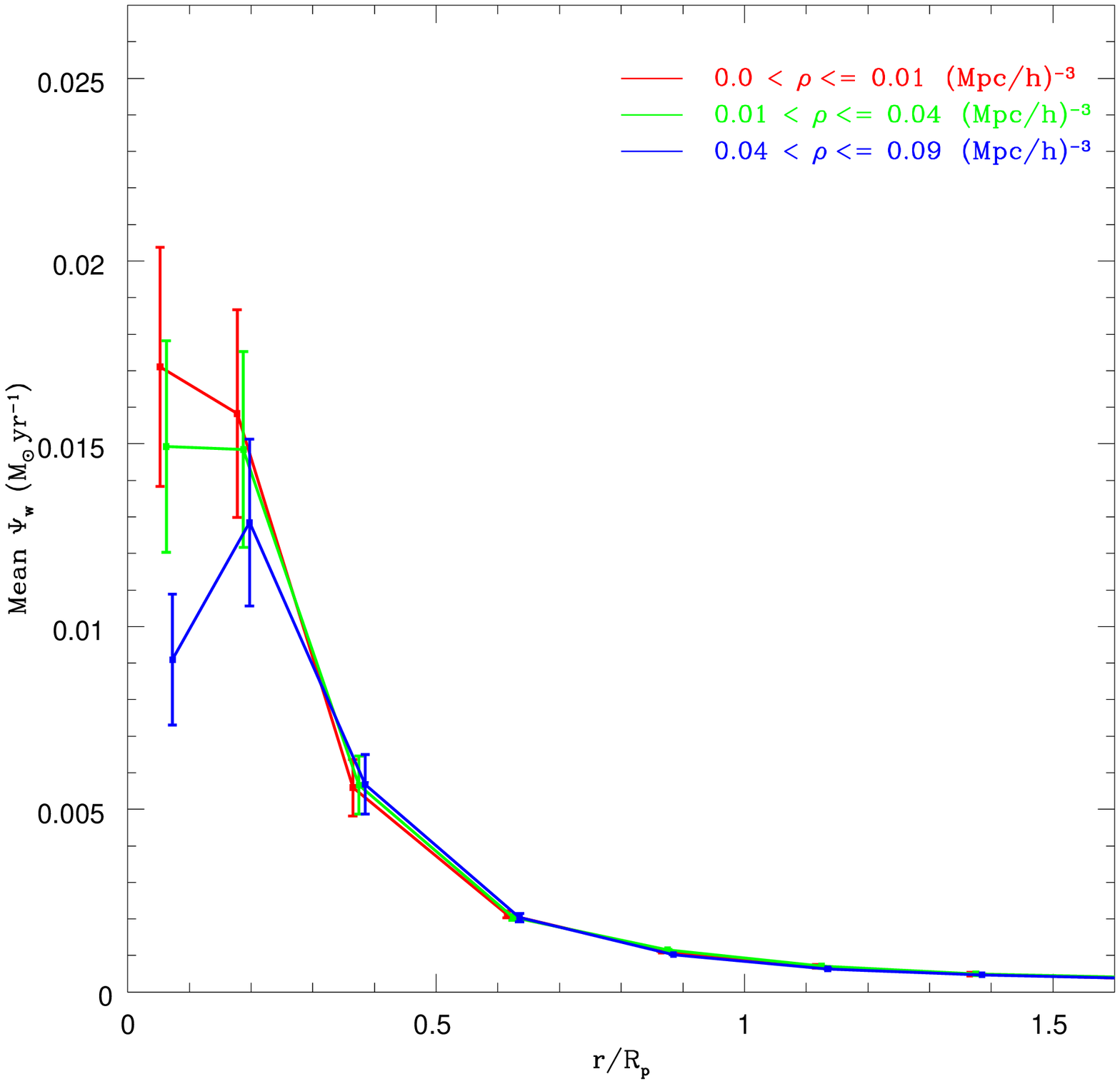}
}
}
\caption{Mean of the distribution of weighted mean SFRs $\Psi_{w}$ ($M_{\odot} \,yr^{-1} $) within successive radial annuli as a function of the local galaxy density $\rho$. Top left panel: For the highest SF early-type galaxies. Top right: For highest-SF late-type galaxies. Bottom left panel: For all early-types in the full sample (no cuts in the total SFR are made). Top right panel: For all late-type galaxies in the full sample.   
\label{fig:radialplots-mean}}
\end{figure}

Let $\Psi(r,\rho)_E$ be the mean SFR profile of early-types at density
$\rho$, $\Psi(r,\rho)_L$ the mean SFR profile of late-types at density
$\rho$ and $\Psi(r,\rho)_T$ the mean SFR profile of all types
at density $\rho$. We will use $\rho_1$, $\rho_2$, $\rho_3$ to refer to
the lowest, intermediate, and highest density environments respectively.
Figure~\ref{fig:radialplots-mean} shows the radial
variation of $\Psi(r,\rho)_E$ and $\Psi(r,\rho)_L$. The high value of
the mean, compared to the 75th percentile, reflects an underlying distribution
of $\psi_w$ that is skewed towards higher values of SFR. The trends in the
mean reflect many of those observed with the 75th percentile in that
there is little effect of environment on the outskirts of the galaxies of
either type, but there is a more marked suppression in the mean of $\Psi$
in the galaxy center in either type. This is true for both the high SF
galaxies and for the full sample of early and late-types (the mean being
more sensitive to the high SFRs in the most active galaxies). For the
high SF late-types, there is a relatively high mean $\Psi$ in the galaxy
center ($r/R_p \le 0.125$) and this is suppressed by about $2\sigma$
in the highest density environment $\rho_3$. In the high-SF early type
galaxies, there is a $2\sigma$ suppression in the first two inner annuli,
up to $r \le 0.25R_p$.

Suppose the suppression of SF is just due to the density-morphology relation,
simply a higher fraction of early-types in high-density environments. We
can use the individual SF profiles of early and late-type galaxies at
low densities $\rho_1$ to determine what the profile for the composite
sample would be at the highest densities $\rho_3$. To do this we
take $\Psi(r,\rho_1)$ for early and late-types, explicitly
assuming these remain the same within different environments,
and average them, weighted by the relative proportion of early and late types
at $\rho=\rho_3$. This produces an artificial composite profile at 
$\rho=\rho_3$ that reflects the profile expected if it arose solely from the
density-morphology relation:
\begin{equation}
\label{eq:artprofile}
\Psi(r,\rho_3)_{\rm artificial} = \frac{N_E}{N_E + N_L}(\rho_3) \times \Psi(r,\rho_1)_E + \frac{N_L}{(N_E + N_L)}(\rho_3) \times \Psi(r,\rho_1)_L.
\end{equation}
We compare this artificial profile at $\rho=\rho_3$ to the one
actually observed at $\rho=\rho_3$:
\begin{equation}
\label{eq:obsprofile}
\Psi(r,\rho_3)_{\rm observed} = \frac{N_E}{N_E + N_L}(\rho_3) \times \Psi(r,\rho_3)_E + \frac{N_L}{(N_E + N_L)}(\rho_3) \times \Psi(r,\rho_3)_L.
\end{equation}
$\Psi(r,\rho_3)_{\rm observed}$ is of course simply equal to
$\Psi(r,\rho_3)_T$.

\begin{figure}
\centerline{\rotatebox{0}{
\includegraphics[width=.55\textwidth,height=0.5\textwidth]{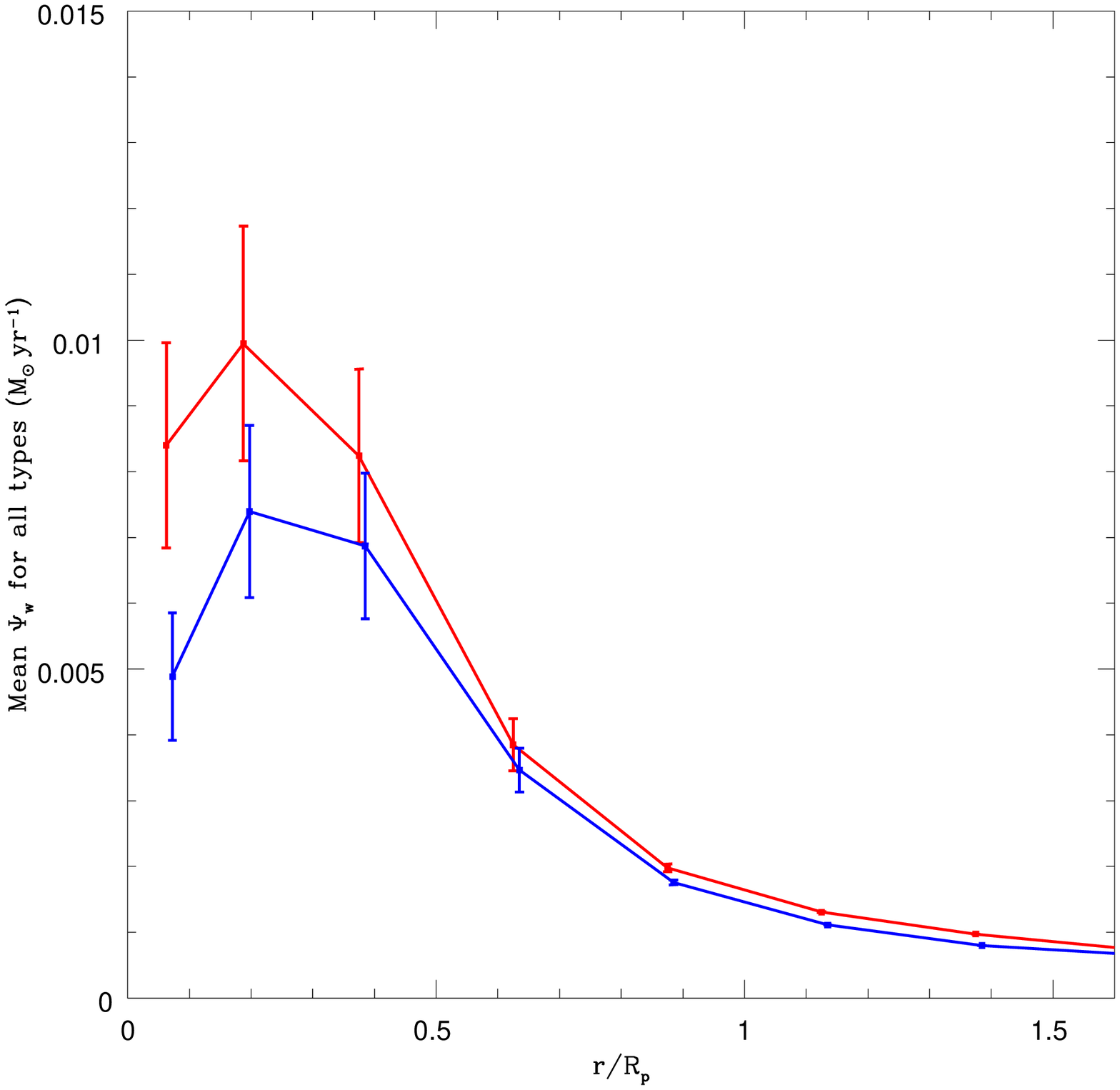}
}
}
\vspace{0.3cm}
\centerline{\rotatebox{0}{
\includegraphics[width=.55\textwidth,height=0.5\textwidth]{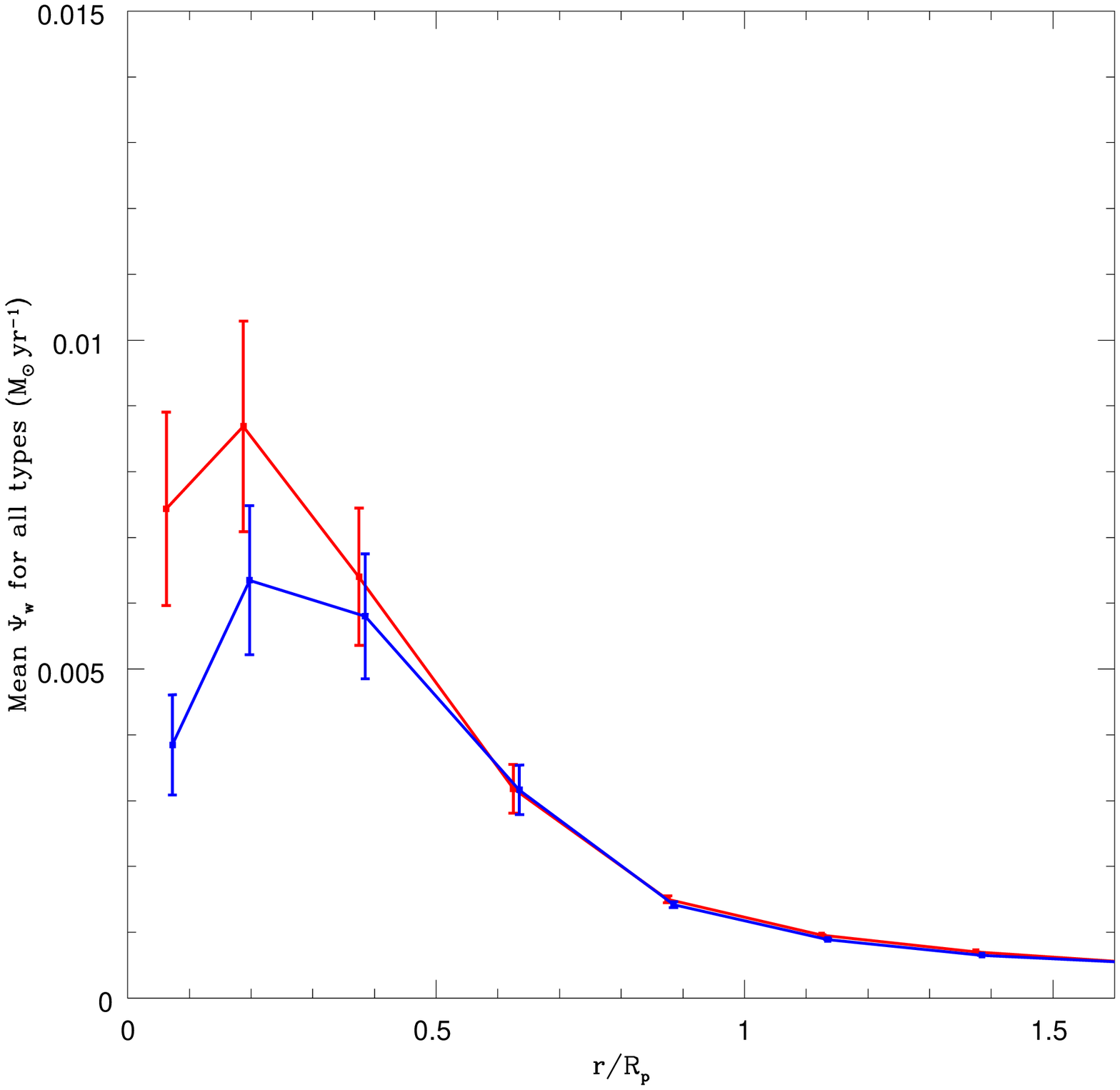}
}
}
\caption{The contribution of the density-morphology relation to the observed dependence of SFR on galaxy density. The red line is the artificial composite SFR radial profile (for both early and late-types) in the highest density interval for the highest SF galaxies. The profile is obtained by taking the SFR profiles of the highest SF early and late-types in the lowest density interval, averaging them and weighting by the relative proportion of early and late-types in the highest density environments (see \S\ref{subsec:density_morph}). This artificial composite profile thus predicts what the SFR profile would look like in the highest density environment if it arose from the density-morphology relation alone. This profile can then be compared to the observed composite profile in the highest density environment (in blue). The bottom panel shows the composite artificial and observed SFR profiles for the full sample of galaxies (not just the highly star-forming ones).
\label{fig:density-downsizing-test}}
\end{figure}

The result of this composite SFR profile is shown in
Figure~\ref{fig:density-downsizing-test} for both the highest SF
galaxies and for the full sample. The observed SFR profile in both cases
is significantly below the artificial SFR profile, particularly in the
innermost annulus $r \le 0.125 R_p$. {\em This shows directly that the
density-morphology relation alone cannot give rise to the observed suppression
in SF in centers of galaxies in high density environments.} This answers
the third of the questions we initially posed, but raises another.
Since the effect is not due to the morphology-density relationship alone,
what is the mechanism or mechanisms that drive the observed suppression in SF?
Before addressing this, we briefly investigate the possibility that our
results, which are a consequence of the highest SF galaxies,
are a simple consequence of higher SFRs in more luminous galaxies.

\subsection{The Effect of Luminosity
\label{subsec:luminosity}}

\begin{figure}
\centerline{\rotatebox{0}{
\includegraphics[width=.55\textwidth,height=0.55\textwidth]{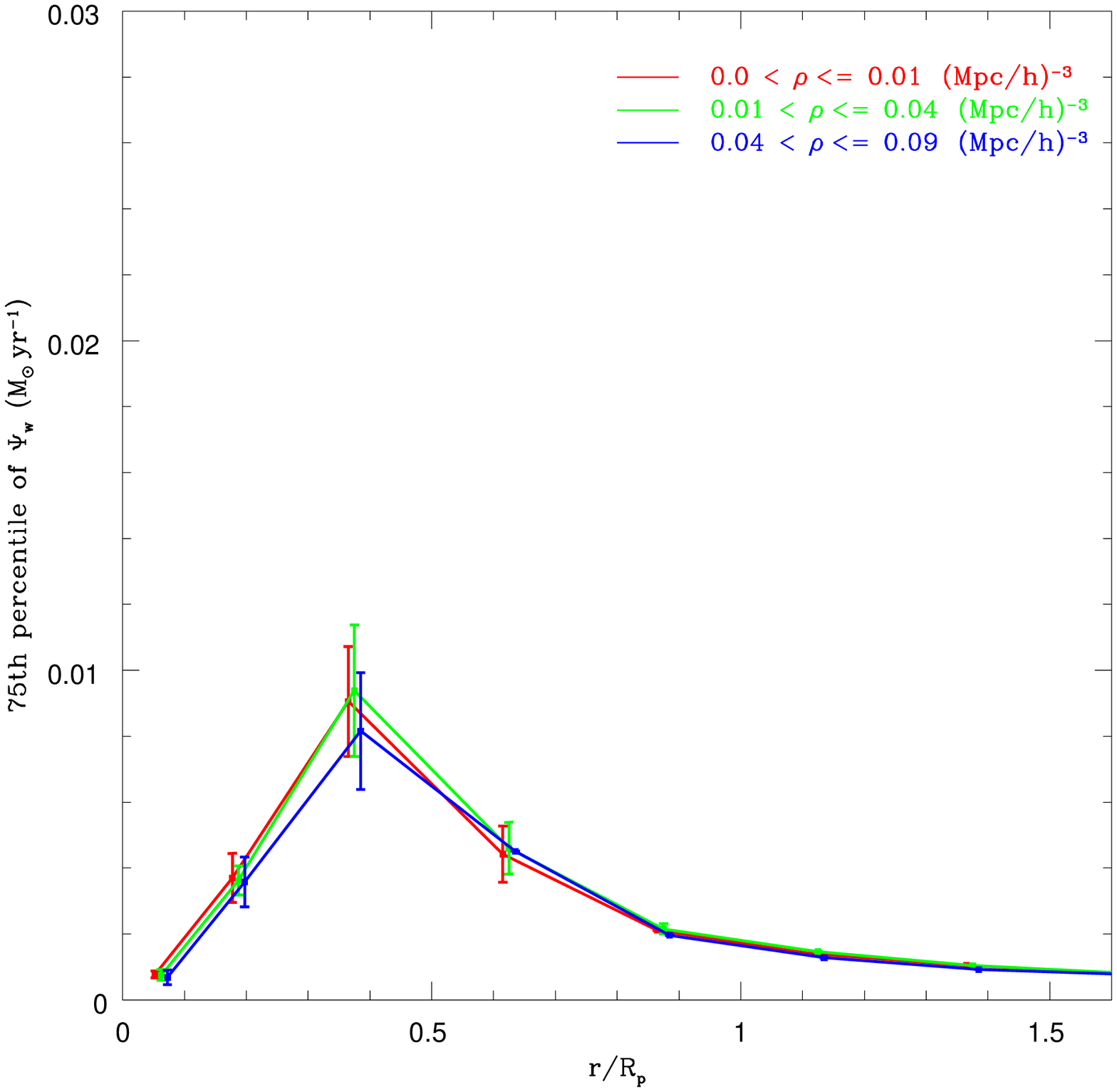}
\hspace{1.0cm}
\includegraphics[width=.55\textwidth,height=0.55\textwidth]{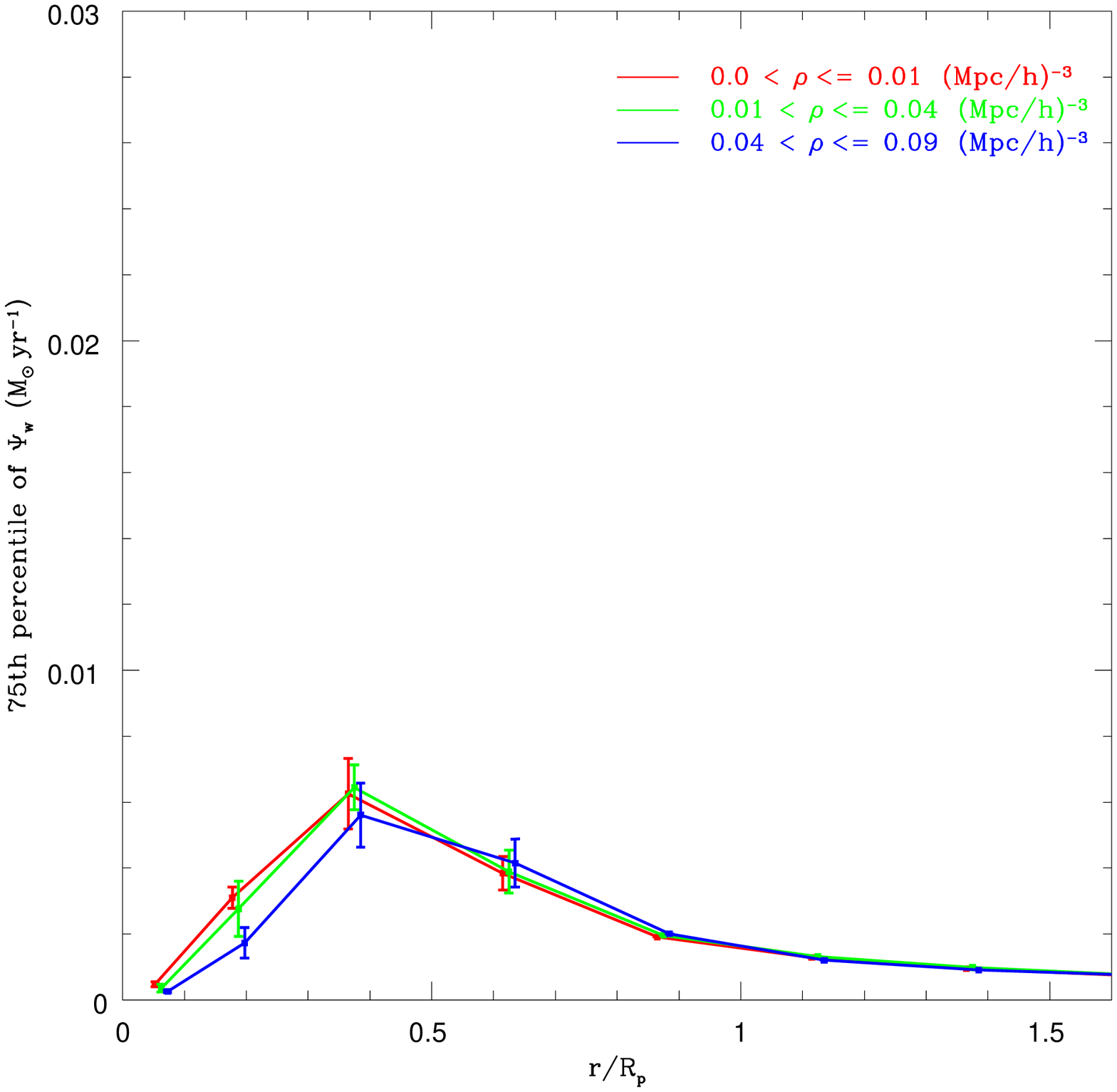}
}
}
\vspace{0.4cm}
\centerline{\rotatebox{0}{
\includegraphics[width=.55\textwidth,height=0.55\textwidth]{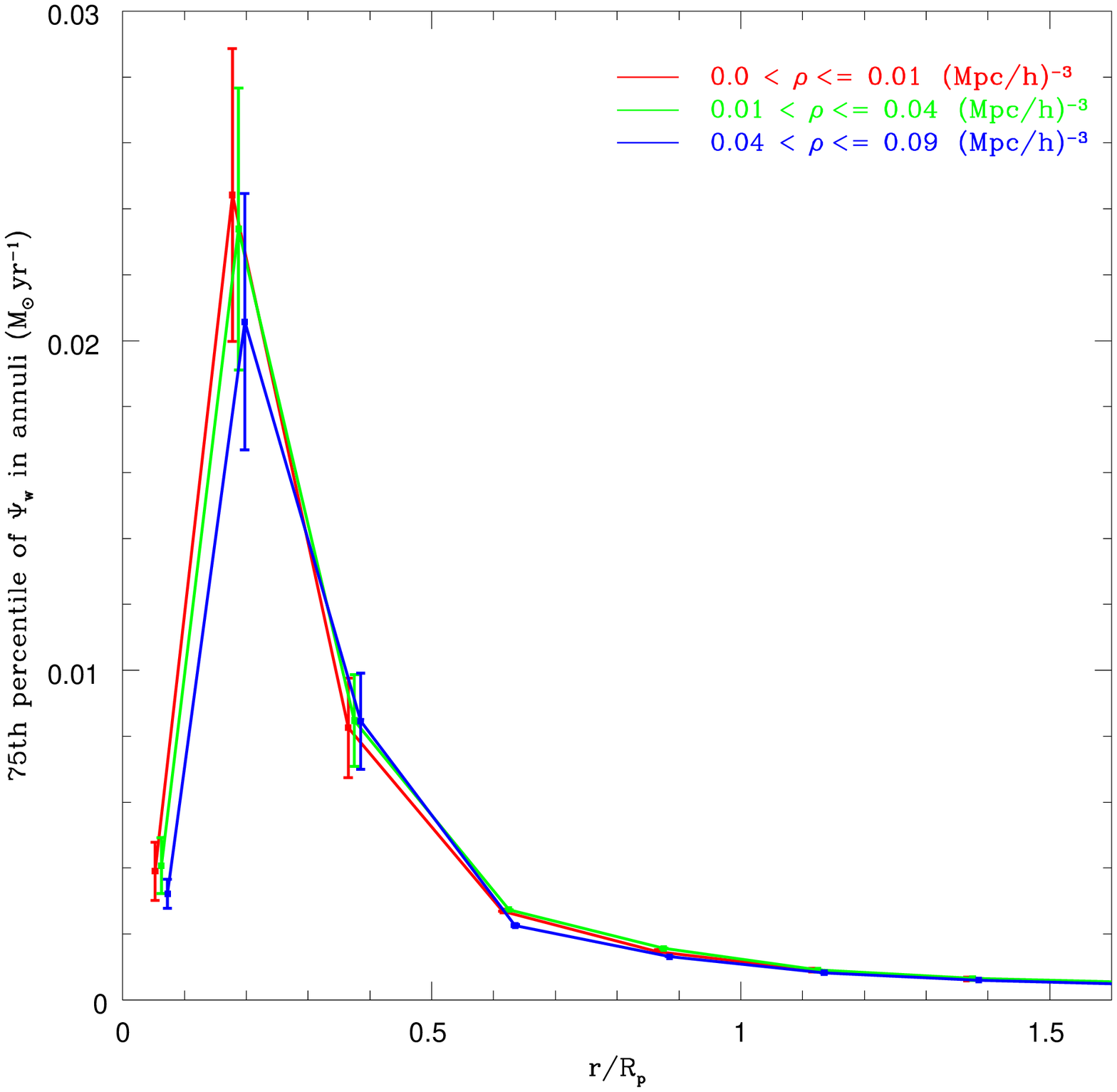}
\hspace{1.0cm}
\includegraphics[width=.55\textwidth,height=0.55\textwidth]{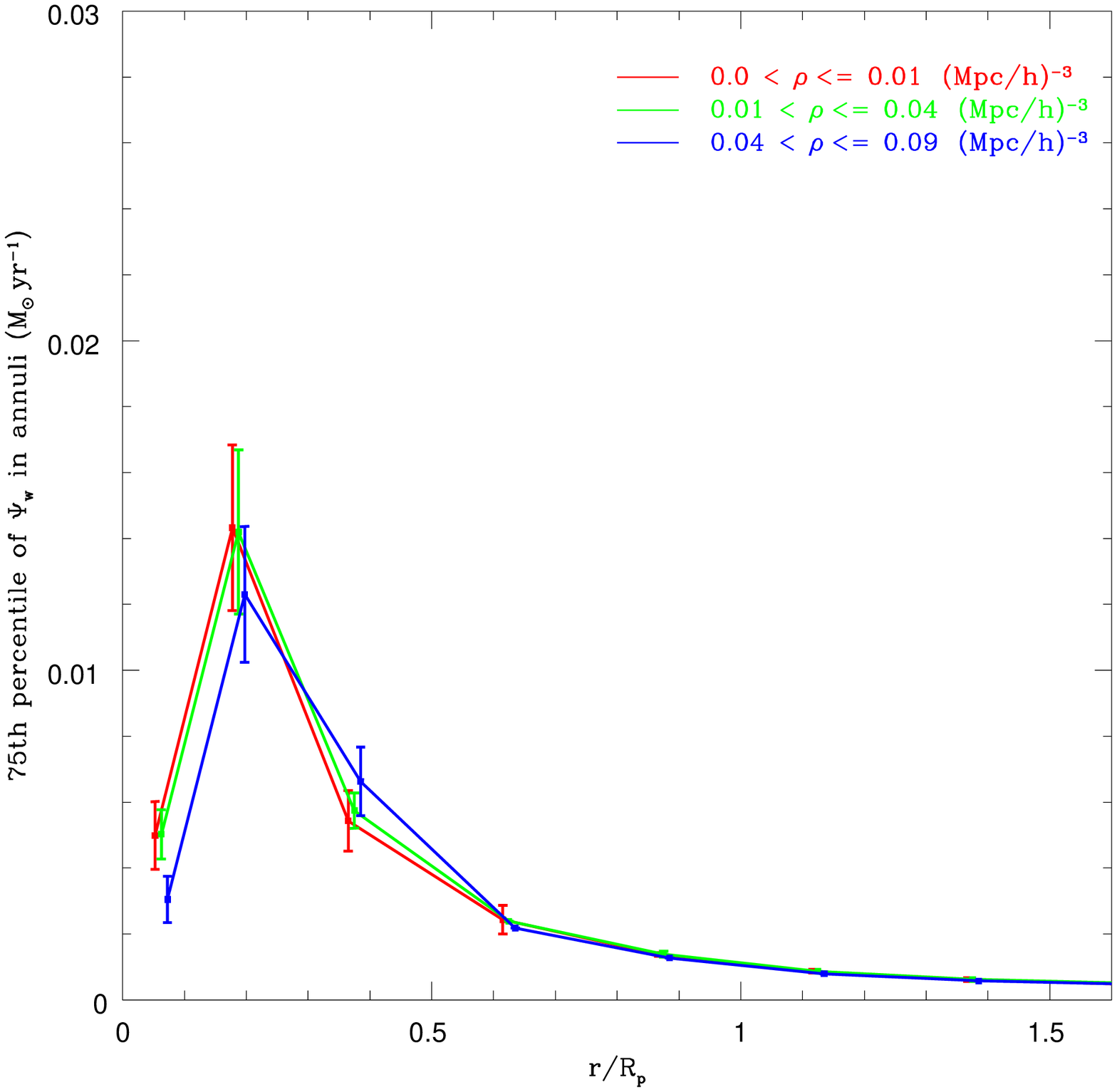}
}
}
\caption{Top panel: 75th percentile of the distribution of weighted mean SFRs $\Psi_{w}$ ($M_{\odot} \,yr^{-1} $) within successive radial annuli as a function of the local galaxy density $\rho$ for early-type galaxies in our sample in intervals of absolute magnitude: $-22.5 <M_r \le -21.5$ (top left) and $-21.5 <M_r \le -20.5$ (top right). Bottom panel: for late-type galaxies for the same two absolute magnitude intervals.
\label{fig:typelumplots}}
\end{figure}

Here we investigate the effect of galaxy luminosity on the radial variation
of SF and its dependence on environment. \citet{Par:07} studied the color
gradients of galaxies brighter than $M_r=-18.5$ as a function of the
local galaxy density in the SDSS. For early-type galaxies, they found
no environmental dependence of the color gradient at a given absolute
magnitude, and in addition found that the gradient is almost independent
of the luminosity as well. They also found that for late-type galaxies
which are bright, there is no dependence of the gradient on environment,
while for fainter late-types, there is a weak dependence on environment -
fainter galaxies were seen to become bluer at the outskirts (relative to
the galaxy center) at low densities while the color gradient vanishes in
high density environments. In order to determine if the luminosity of
galaxies has some effect on the radial variation of SF and the way it
depends on the galaxy environment, we split our galaxies into narrower
intervals of absolute magnitude and examine the radial distribution of
SF within the early and late-type galaxies in each interval, across the
same range of densities. K-corrected absolute magnitudes are obtained from the CMU-Pitt SDSS Value Added Catalog (VAC) database\footnote{http://nvogre.phyast.pitt.edu/dr4/}, where the k-corrections had been applied according to \citet{Bla:03}. The top panel of Figure~\ref{fig:typelumplots} shows the SFR profiles (75th percentile of $\psi_w$) for early-type galaxies in
the absolute magnitude intervals $-22.5<M_r\le -21.5$ and $-21.5<M_r\le -20.5$
respectively, and the bottom panel is for the late-type galaxies for
the same absolute magnitude intervals.
The general characteristics of the profiles for both types remain unchanged
when we split our sample into these narrower intervals of luminosity.
We detect no effect of the environment on the SF in the brighter sample but
do recover the suppression of SF with increasing local density in 
the fainter sample.

For the brighter early-type galaxies we find no
significant change in the variation of SF with galaxy density in
either the center or outskirts of the galaxies, which is consistent with
the findings of \citet{Par:07}. But for the fainter early-type galaxies
we do detect a suppression of SF at higher densities. There is a decrease of
$2.5\sigma$ in $0.125 < r/R_p \le 0.25$ between the lowest and highest
local density intervals, while there is no change in the SFR with
increasing density in the outskirts.

For the late-type galaxies, just as for the early-types, there is no
change in the mean SFR at any radius with environment for the brighter
sample, which is consistent with the \citet{Par:07} result. For the
fainter sample, there is a decrease in the mean SFR (approximately a $2\sigma$
difference) in the region $r/R_p \le 0.125$, while the mean SF in the
outskirts is again unchanged. \citet{Par:07}, however, finds that at the highest
densities, both the center and the outskirts become redder (giving zero
color gradient). 

There are some conclusions to be drawn here. First, there is no indication that our
primary results are a direct consequence of a SFR-luminosity relation, since
it is the fainter subsample that contributes primarily to the trends with environment that are observed.
There remains the caveat that the sample selection was very different between the two groups: \citet{Par:07} classified galaxies based on their locations in the {\em u-r} versus {\em g-i} color gradient space as well as concentration index space \citep{Par:05} whereas we use concentration alone. 
Also, degeneracies, such as that between age and metallicity, may contribute to the difference between our result and that of \cite{Par:05}. Finally, the disparity between the two sets of results may also be explained to some extent by the inclusion of dust
obscuration in the pixel-z analysis. SFR gradients may be observed in
the absence of color gradients, as the effect of dust will be to redden
otherwise bluer star forming stellar populations.

\section{Discussion 
\label{sec:discussion}}

The distinct radial SF profiles separate the two galaxy types cleanly.
Early-types have a lower mean SFR throughout compared to late-type
galaxies. The environment is found to affect the highest SF early-type
galaxies only while affecting the late-type galaxies in both the highest
and second highest quartiles of their total SFR distribution. In the
highest SF early-type galaxies, the suppression of SF takes place in
$0.125 < r/R_p \le 0.25$ while in the the late-type galaxies, the
suppression takes place at $r/R_p \le 0.125$. This suggests that the
suppression in SF in these active galaxies of either type, is independent
of the established density-morphology relation. 

It is worth noting that, as found for the full sample in Paper I, the
outskirts of galaxies in either the early or late-type galaxies are not
affected by a changing environment. This implies that the outskirts of
either type of galaxy are not significantly affected by processes such
as ram-pressure stripping and galaxy harassment or interaction between
galaxies, processes which remove cold gas available for star formation,
and which should primarily affect the outskirts of galaxies before they
affect the inner regions. The lack of effect in the outskirts could alternatively be due to the fact that these processes affect only a small fraction of the population. Our results are consistent with the conclusions of \cite{Coo:06} who studied the relationship between galaxy properties and environment at $0.75 < z < 1.35$ in the DEEP2 Galaxy Redshift Survey. Their findings suggest that cluster-specific processes, such as ram-pressure stripping and harassment, are not needed to establish the color-density relation and that the group environment too plays a critical role in quenching star formation. A similar result is echoed at low and intermedidate redshifts \citep{vandenBosch:08,Pat:08}. 

The short timescales for the physical mechanisms of SFR quenching would also make them difficult for our method to detect as they will only be detectable for a small fraction of the
sample at any given redshift, and this will be hidden by our quartile
sampling statistics. This is supported by the results of \citet{DD:06},
who find a reduction in the {\em number\/} of galaxies with neutral hydrogen
(HI) in high-density environments, but no significant trend with
environment in the star formation rate or efficiency of star formation in
HI galaxies. There is also the possibility that some of these
processes could be taking place at much lower densities than we are
probing.

The results of \S\,\ref{subsec:density_morph} suggest that the suppression
of SF cannot be due to the density-morphology relation alone. Given that
both physical ``infall and quench'' mechanisms and the density-morphology
relation are ruled out as the main mechanisms of SF suppression in either
type of galaxy, we explain the trends observed by extending the
``downsizing of SF'' hypothesis laid out in Paper I in terms of galaxy
morphology.

Downsizing is characterized by a decrease in the mass of galaxies that
dominate the SFR density with increasing cosmic time. This was first
suggested by \cite{Cow:96} who found that the maximum rest-frame K-band
luminosity of galaxies undergoing rapid star formation has been decreasing
smoothly with time in the redshift range $z=0.2-1.7$. This is supported
by more recent studies of star-formation histories of galaxies in both
the local and distant universe. \citet{Hea:04} observed that the most
massive local galaxies seen in the SDSS also appear to be dominated by
stars which formed at early epochs. \citet{Jun:05} studied the cosmic
SFR and its dependence on galaxy stellar mass in galaxies in the Gemini
Deep Deep Survey (GDDS) and found that the SFR in the most massive
galaxies ($M_\star > 10^{10.8} M_\odot$) was 6 times higher at $z=2$ than at
present and that the SFR at $z=2$ falls sharply to reach its present ($z=0$) value
by $z \approx 1$. \citet{Pan:04} found no evolution in the stellar mass
function of galaxies in the SDSS in the redshift range $0.05 \le z \le
0.34$ indicating that almost all stars were formed by $z \approx 0.34$
with little SF activity since then. In a radio-selected survey, \cite{Sey:08} also determined that high mass galaxies only contribute significantly to the SFR density of the universe at high redshifts, with low-mass systems dominating at lower redshift.

To interpret our results we consider a population of galaxies
that is active and highly star forming at high redshifts. These will
ultimately form early-type galaxies and include in particular the most
massive systems that form within the more massive dark-matter halos in
dense environments. As the SF, in the ``downsizing" scenario, is progressively
associated with less massive galaxies in lower-mass halos and less dense
environments, the SF moves from being dominant in cluster regions at
high redshift to being dominant in low-density regions at low redshifts.
This explains the density-morphology relation. To explain the radial
dependence of SF with environment -- the central suppression of SF in
the high-SF galaxies -- we consider that bulges within late-type
galaxies may form and evolve similarly to early-type galaxies \citep{Dri:07}.
This will result in late-type galaxies in high-density environments having
bulges that formed the bulk of their stars earlier than similar late-types
in lower-density environments. These will consequently show the same kind
of reduced current SF, purely as a consequence of their rapid early
evolution, compared to late-types in lower-density environments.
This is also consistent with the observed reduction in the SFR of
early-types in high-density environments.

A natural consequence of downsizing in galaxies would be that the
SFR-density relation should invert at higher redshifts, with increased
SFR in the more dense environments. This has been measured
by \citet{Elb:07} using data from the Great Observatories Origins Deep Survey
(GOODS) at $z \approx 1$. They found the SFR-density relation observed
locally was reversed at $z \approx 1$, with the average SFR of galaxies
increasing with increasing local density. \cite{Coo:08}, using galaxy samples drawn from the SDSS and the DEEP2 Galaxy Redshift Survey, found an inversion of the SFR-density relation from $z \approx 1$ to $z \approx 0$. They found that this evolution in the SFR-density relation is driven, in part, by a population of bright, blue galaxies in dense environments at $z \approx 1$ \citep{Coo:06}. This population is thought to evolve into members of the red sequence from $z \approx 1$ to $z \approx 0$.
 This adds additional support to the interpretation of our observed SFR suppression at low redshifts in terms of downsizing.

Reproducing the radial SFR profiles and their density dependence may require
some addition to current models. It has recently been shown \citep{Nei:06}
that downsizing could be a natural outcome of hierarchical structure formation
if mass assembly and star formation are treated as distinct processes that
proceed in opposite directions, and if a characteristic mass for SF truncation
is introduced. Stars can form first in the small building blocks of today's
massive galaxies. If gas processes limit galaxy formation to dark matter
halos above a minimum mass, a certain downsizing arises naturally
from the mass assembly process itself \citep{Nei:06}. \citet{Cat:08}
who studied the origin of downsizing of elliptical galaxies using the
mean stellar ages of galaxies, showed that this could result naturally
from a shutdown of star formation in dark matter halos above a critical
mass of $10^{12} M_\odot$. Above this mass there is stable shock heating
which truncates the star formation.

Our observations could well be driven by the downsizing of SF together
with a treatment of late-type galaxy bulges in a similar fashion
to early-type bulges, resulting in the observed central suppression of
SF in high density environments. More work needs to be done on the theoretical
front, however, if the observed dependence of the radial SF on the environment
is to be reproduced in the models.

\section{Limitations of this analysis
\label{sec:futurework}}

In future work, we would like to perform our analysis with additional morphological proxies. An improvement in the accuracy of the classification using the concentration index $C_{in}$ alone could potentially be achieved by calculating {\tt petroR50\_r} and {\tt petroR90\_r} using elliptical rather than circular apertures. This would address the problem of late-type galaxies with small axis ratios (edge-on) being classified as early-types erroneously. This is unlikely to change our statistical results significantly since the random orientation of galaxies in our sample, both on the sky and to the line of sight, will act to mitigate any systematic effects when we average over annuli. This is supported by the results of selecting only face-on and edge-on galaxies of either type in our sample using the axis-ratios given in the CAS and repeating our analysis - the 75th percentiles for $\Psi$ are found to be invariant to the effect of inclination. More recently, a catalog of visually-classified morphologies of galaxies in the SDSS has become available through the Galaxy Zoo project \citep{Lin:08}. Since this avoids biases introduced by other morphological proxies such as concentration or color, we would aim to repeat our analysis using this new classification. 

The second issue we would like to attempt in future work is to implement a full seeing correction to the colors in the pixels. More than half of the galaxies in our sample have innermost radial annuli ($r < 0.125 R_p$) which are larger in angular size than the PSF width. We have chosen a rather coarse grid of stellar population parameters (age, e-folding time, dust obscuration and metallicity). Color corrections due to seeing variablity across different passbands are not expected to change the values of the inferred parameters of the stellar populations in the pixels beyond the bounds of the error range that the technique has estimated, although it cannot be ruled out for a small number of pixels where degeneracies between the stellar population parameters are severe. In general, the inferred parameters for these severe cases should be associated with large uncertainties but this may not always be the case. In future work, we will attempt to quantify the effect of these degeneracies and color corrections on the estimated values of the inferred parameters and their associated uncertainties. Along with this, we also aim to implement a full seeing model for the pixels and a correction for the seeing variability across the various passbands.

\section{Conclusions
\label{sec:conclusions}}

We use the ``pixel-z'' technique as in Paper I, here dividing SDSS galaxies by
morphology in order to study the role of the density-morphology relation
on the spatially resolved SF in galaxies. A density-morphology relation is
quantified for our volume-limited galaxy sample. We find that a
suppression of SF occurs in the most active and highly star-forming systems
in the highest density environments for both broad galaxy morphological types.
We find that neither ``infall and quench'' mechanisms nor the
density-morphology relation by themselves can account for this observed
suppression of SF in either early or late-type galaxies. The results
are consistent with the picture of ``downsizing'' in galaxy formation,
together with the idea that late-type galaxy bulges form and evolve
in a similar fashion to early-type galaxies, leading to a lower SFR
in high-density environments. 

In particular, we observe the following:
\begin{itemize}
\item Early and late-type galaxies in the SDSS each have a distinct spatial
variation of SF, with early types having a SFR distribution that extends
further (relative to the galaxy scale length) compared to late-type galaxies.
\item A suppression of SF occurs in the most active and highly
star-forming systems in the highest density environments for either
galaxy type. The suppression takes place in the innermost regions of the
galaxies, occurring at $0.125<r/R_p\le 0.25$ for early-type
galaxies and $r/R_p\le 0.125$ for late-type galaxies.
\item In early-type galaxies, only those in the top quartile of the total
SFR distribution ($ \rm{SFR}>0.45 \,M_{\odot} \, \rm{yr^{-1}} $) show
any significant SF suppression in cluster environments. In late-types galaxies,
there is a much larger range of total SFR in these galaxies (the top two
quartiles) where a suppression of SF is observed.
\item We find no significant environmental dependence when considering
the full sample of early and late-type galaxies, indicating that the
trends with environment are driven by the highly star-forming galaxies.
\item The suppression of SF is seen primarily in lower-luminosity galaxies
($-21.5<M_r\le -20.5$) in our sample, while we detect no environmental
dependence on the radial distribution of SF for brighter galaxies.
\item By appropriately weighting the average SFR profiles of early and
late-type galaxies in low-density regions by the proportions of these types
found in high-densities, we show that the density-morphology relation alone
cannot account for the suppression of SF in the highest density environments.
\end{itemize}

In future work we will probe whether the ``downsizing'' of star formation
in the centers of galaxies is indeed responsible for our observed trends by exploring
galaxy populations as a function of stellar mass, together with the
local density. Since ``downsizing'' is concerned with the mass dependence
of the SFR history of galaxies, this will allow a more detailed exploration
of the origins and likely evolution of the observed variation in
the spatially resolved SF in galaxies.

\section{Appendix A: Calculation of the flux in the pixels 
\label{sec:appendix}}

We utilize one of the photometric data products of the SDSS: the fpAtlas images as described in Paper I. These are cutouts from the imaging data in all five bands ({\em u,g,r,i,z}) of all detected objects in the survey. The photometry in every pixel that belongs to each galaxy of the fpAtlas image (in each passband) needs to be calibrated in order to obtain a measured flux from the counts in that pixel. In this section, we illustrate the steps that need to be taken to do this. The photometric calibration of the Atlas images was done according to the {\em asinh} magnitude system developed by \cite{Lup:99} and foreground Galactic extinction is also corrected for. We follow the prescription for the photometric calibration of the imaging data that is given in the SDSS Photometric Flux Calibration web page\footnote{http://www.sdss.org/dr4/algorithms/fluxcal.html}. We summarize the basic steps below.

We obtain a count rate $t/t_0$ from the net count $N_{DN}$ (in ``Data Numbers'', $DN$) in each pixel in each of the passbands ({\em u,g,r,i,z}).  

\begin{equation}
\frac{t}{t_0} = \frac{N_{DN} \times 10^{0.4 \times (a_0 + (k \times A))} }{T}
\end{equation}

where $t/t_0$ is defined to be the count rate and $t_0$ is the zero point count rate and is given by $t_0 = 10^{-0.4 \times a_0}$. $T$ is the exposure time of the SDSS 2.5 meter telescope, $k$ is the atmospheric extinction coefficient in a particular passband and $A$ is the airmass (optical path length relative to zenith for light travelling through the Earth's atmosphere) for that passband. $a_0$ is the zero-point in each passband. $k$, $A$ and $a_0$ are obtained from the {\tt Field} table of the SDSS Catalog Archive Server (CAS, http://cas.sdss.org/dr4/). For the {\em r'} band, these correspond to the {\tt kk\_r}, {\tt airmass\_r} and {\tt aa\_r} columns.  

The count rate is then converted to an SDSS {\em asinh} magnitude $m$ (and thence to an AB flux) using:

\begin{equation}
m = -\frac{2.5 \times (asinh((t/t_0)/2b) + ln(b))}{ln(10)}
\end{equation}

where b is a softening parameter for a particular photometric band ({\tt bPrime\_r} for the {\em r'} band in CAS). The magnitudes are corrected for foreground Galactic extinction using the values of extinction (in magnitudes) in each band that are found in the {\tt PhotoObjAll} table ({\tt extinction\_r} for the the {\em r'} filter). These values are obtained from the maps of foreground dust infrared emission produced by \cite{Sch:98}.



The error in the counts $N_{DN}$ is calculated from the Poisson error from the photoelectrons which are counted by the CCD detectors, and the total noise contributed by read noise and dark currents:

\begin{equation}
  \delta N_{DN} =  \sqrt{(\frac{N_{DN}+ N_{sky}}{G}) + N_{pix} \times (D + \delta N_{sky})}
\end{equation}

where $N_{sky}$ is the number of sky counts (in $DN$) over the area considered, obtained from the average sky background level in the frame (in $maggies/arcsec^2$) that the object was found in, given in the {\tt sky\_r} column in the {\tt Field} table for the {\em r'} band. $N_{pix}$ is the area covered by the object in pixels ($N_{pix} = 1$ for one pixel). $D$ is the noise due to dark variance (obtained from the {\tt darkVariance\_r} column in the {\tt Field} table for the {\em r'} band) and $\delta N_{sky}$ is the error on the estimate of the average sky level in the frame (obtained from the {\tt skyErr} column in the {\tt Field} table and then converting $maggies/arcsec^2$ to $DN/pix$). 

Finally, we work out the error in the SDSS magnitude (from which we can derive an error in the flux):

\begin{equation}
\delta m = \frac{2.5 \times \delta N_{DN} \times  10^{0.4 \times (a_0 + k \times A)} }{2bT \times ln(10) \times \sqrt{1 + (\frac{t}{2bt_0})^2} }
\end{equation}

The Point Spread Function (PSF) is not taken into account in our analysis. The median seeing value reported in DR4 was $1.4''$ so the PSF will be $3-4$ pixels across typically. We have ignored this effect for now
since, as long as the galaxies are sufficiently resolved, there will
be useful information to be gained regardless of whether the small-scale
details of structure that will be smeared by the PSF can be extracted. Although seeing can indeed affect the innermost annulus in galaxies, for over 53\% of galaxies, this annulus ($r \le 0.125R_p$) is larger in angular scale than the PSF width. 

\section{Acknowledgments
\label{sec:acknowledge}}

We would like to thank the referee for their valuable input which has improved this paper.
NW would like to thank Andrew Zentner, Jeff Newman and Michael Cooper for useful discussions. This material is based upon work supported by the National Science Foundation under Grant No. 0851007 and AST 0806367. AMH acknowledges support provided by the Australian Research Council in the form of a QEII Fellowship (DP0557850). This material is also based upon work supported by the National Science Foundation under the following NSF programs: Partnerships for Advanced Computational Infrastructure, Distributed Terascale Facility (DTF) and Terascale Extensions: Enhancements to the Extensible Terascale Facility.

Funding for the SDSS and SDSS-II has been provided by the Alfred P. Sloan Foundation, the Participating Institutions, the National Science Foundation, the U.S. Department of Energy, the National Aeronautics and Space Administration, the Japanese Monbukagakusho, the Max Planck Society, and the Higher Education Funding Council for England. The SDSS Web Site is http://www.sdss.org/. The SDSS is managed by the Astrophysical Research Consortium for the Participating Institutions. The Participating Institutions are the American Museum of Natural History, Astrophysical Institute Potsdam, University of Basel, University of Cambridge, Case Western Reserve University, University of Chicago, Drexel University, Fermilab, the Institute for Advanced Study, the Japan Participation Group, Johns Hopkins University, the Joint Institute for Nuclear Astrophysics, the Kavli Institute for Particle Astrophysics and Cosmology, the Korean Scientist Group, the Chinese Academy of Sciences (LAMOST), Los Alamos National Laboratory, the Max-Planck-Institute for Astronomy (MPIA), the Max-Planck-Institute for Astrophysics (MPA), New Mexico State University, Ohio State University, University of Pittsburgh, University of Portsmouth, Princeton University, the United States Naval Observatory, and the University of Washington. 

\acknowledgments


\end{document}